\documentclass[useAMS,usenatbib,useasmath]{mn2e}
\usepackage{epsfig}
\usepackage{longtable}
\usepackage{times}
\usepackage{amsmath}
\usepackage[usenames,dvipsnames]{color}

\newcommand{\iue}{\emph{IUE}}
\newcommand{\stis}{\emph{STIS}}
\newcommand{\cosp}{\emph{COS}}
\newcommand{\ghrs}{\emph{GHRS}}
\newcommand{\fos}{\emph{FOS}}

\newcommand{\cmfgen}{{\sc CMFGEN}}
\newcommand{\tlusty}{{\sc TLUSTY}}
\newcommand{\fastwind}{{\sc FASTWIND}}
\newcommand{\xshooter}{\emph{X-shooter}}

\newcommand{\mdot}{\hbox{$\dot M$}}

\newcommand{\zsun}{\hbox{$Z_\odot$}}

\newcommand{\msun}{\hbox{$M_\odot$}}
 \newcommand{\msunyr}{\hbox{$M_\odot\,$yr$^{-1}$}}
\newcommand{\vinf}{\hbox{$V_\infty$}}
\newcommand{\msp}{$M_{\rm spec}$}
\newcommand{\xit}{$\xi_{t}$}
\newcommand{\vsini}{$v\sin\,i$}

\newcommand{\dmom}{\hbox{$D_{mom}$}}

\newcommand{\teff}{\hbox{$T_{\hbox{\small eff}}$}}

\newcommand{\kms}{\hbox{km$\,$s$^{-1}$}}

\def\lesssim{\mathrel{\hbox{\rlap{\hbox{\lower4pt\hbox{$\sim$}}}\hbox{$<$}}}}
\def\gtrsim{\mathrel{\hbox{\rlap{\hbox{\lower4pt\hbox{$\sim$}}}\hbox{$>$}}}}

\newcommand{\logg}{\hbox{$\log g$}}

\newcommand{\finf}{\hbox{$f_{\infty}$}}
\newcommand{\logL}{\hbox{$\log L/L_{\odot}$}}

\newcommand{\he}{\mbox{He}}

\newcommand{\fe}{\mbox{Fe}}
\newcommand{\si}{\mbox{Si}}

\newcommand{\feiii}{\mbox{Fe~{\sc iii}}}
\newcommand{\feiv}{\mbox{Fe~{\sc iv}}}
\newcommand{\fev}{\mbox{Fe~{\sc v}}}
\newcommand{\fevi}{\mbox{Fe~{\sc vi}}}
\newcommand{\ciii}{\mbox{C~{\sc iii}}}
\newcommand{\civ}{\mbox{C~{\sc iv}}}

\newcommand{\pv}{\mbox{P~{\sc v}}}
\newcommand{\sv}{\mbox{S~{\sc v}}}
\newcommand{\hi}{\mbox{H~{\sc i}}}
\newcommand{\hei}{\mbox{He~{\sc i}}}
\newcommand{\heii}{\mbox{He~{\sc ii}}}
\newcommand{\oii}{\mbox{O~{\sc ii}}}
\newcommand{\oiii}{\mbox{O~{\sc iii}}}
\newcommand{\oiv}{\mbox{O~{\sc iv}}}
\newcommand{\ov}{\mbox{O~{\sc v}}}
\newcommand{\niii}{\mbox{N~{\sc iii}}}
\newcommand{\niv}{\mbox{N~{\sc iv}}}
\newcommand{\siiii}{\mbox{Si~{\sc iii}}}
\newcommand{\siiv}{\mbox{Si~{\sc iv}}}
\newcommand{\nv}{\mbox{N~{\sc v}}}
\newcommand{\lb}{$\lambda$}

\title[]
{No breakdown of the radiatively-driven wind theory in low-metallicity environments\thanks{Based on observations made with the NASA-ESA {\sl Hubble Space Telescope\/} (program GO 12867),
obtained at STScI, which is operated by AURA, Inc., under NASA contract NAS 5-26555.}\thanks{Based on data products from observations made with ESO Telescopes at the La Silla Paranal Observatory under programme ID 085.D-0741}}
 
\author[ J.-C. Bouret et al.]
{J.-C. Bouret $^{1}$\thanks{E-mail: Jean-Claude.Bouret@lam.fr}, T. Lanz$^2$, D. J. Hillier$^3$, F. Martins $^4$, W. L. F. Marcolino$^5$, E. Depagne$^6$ \\
$^1$ Aix Marseille Universit\'e, CNRS, LAM (Laboratoire d'Astrophysique de Marseille) UMR 7326, 13388, Marseille, France\\ 
$^2$ Laboratoire J.-L. Lagrange, UMR 7293, Universit\'e de Nice-Sophia Antipolis, CNRS, Observatoire de la C\^ote d'Azur
        B.P. 4229, 06304 Nice Cedex 4, France\\
$^3$ Department of Physics and Astronomy \& Pittsburgh Particle physics, Astrophysics, and Cosmology Center (PITT PACC), University of Pittsburgh, \\
3941 O'Hara Street, Pittsburgh, PA 15260, USA\\
$^4$ LUPM--UMR5299, Universit\'e Montpellier II \& CNRS, Place Eug\`ene Bataillon, F-34095 Montpellier Cedex 05, France\\
$^5$  Universidade Federal do Rio de Janeiro, Observat\'orio do Valongo. Ladeira Pedro Ant\^onio, 43, CEP 20080-090,
           Rio de Janeiro, Brazil\\
$^6$ South African Astronomical Observatory (SAAO), Observatory Road Observatory Cape Town, WC 7925, South Africa}

\begin{document}

\date{Received 2014 December 18 -- Accepted 2015 February 19}

\pagerange{\pageref{firstpage}--\pageref{lastpage}} \pubyear{2013}

\maketitle

\label{firstpage}

\begin{abstract}
We present a spectroscopic analysis of HST/\cosp\ observations of three massive stars in the low metallicity dwarf galaxies IC 1613 and WLM. 
These stars, were previously observed with VLT/\xshooter\ by \cite{tramper11, tramper14} who claimed that their mass-loss
rates are higher than expected from theoretical predictions for the underlying metallicity. 
A comparison of the FUV spectra with those of stars of similar spectral types/luminosity classes in the Galaxy, and the Magellanic Clouds provides a direct, 
model-independent check of the mass-loss -- metallicity relation
Then, a quantitative spectroscopic analysis is carried out using the NLTE stellar atmosphere code \cmfgen.
We derive the photospheric and wind characteristics, benefiting from a much better sensitivity of the FUV lines to wind properties than H$\alpha$. 
Iron and CNO abundances are measured, providing an independent check of the stellar metallicity.
The spectroscopic analysis indicates that $Z / Z_{\odot} = 1/5$, similar to a SMC-type environment, and higher than usually quoted for IC 1613 and WLM. 
The mass-loss rates are smaller than the empirical ones by \cite{tramper14}, and those predicted by the widely-used 
theoretical recipe by \cite{vink01}. 
On the other hand, we show that the empirical, FUV-based, mass--loss rates are in good agreement with
those derived from mass fluxes computed by \cite{lucy12}.  
We do not concur with \cite{tramper11, tramper14} that there is a breakdown in the mass-loss -- metallicity relation.
\end{abstract}

\begin{keywords}
stars: early-type -- stars: massive -- stars: mass-loss -- stars: wind -- stars: fundamental parameters -- galaxies: individual: IC1613, WLM
\end{keywords}

\section{Introduction}
Hot star winds are radiatively-driven by the transfer of photospheric photon momentum to
the atmospheric material through absorption and scattering by spectral lines \citep{castor75, vink01}. 
The basic properties of stellar winds therefore depend on both the number of metal
lines available to absorb photon momentum and on their ability to absorb (i.e., their optical
thickness). \mbox{CNO} and intermediate elements are dominant line drivers for the outer supersonic part of the
winds, while iron group elements are responsible for the inner, subsonic part where mass-loss
rate is set \citep{vink01}. The mass-loss rates, \mdot, of massive hot stars are therefore expected to depend on
the metal content Z. Based on Monte-Carlo simulations using an extensive line list, \cite{vink01} 
established a theoretical relation, 
$\dot{M} \propto Z^{0.69 \pm 0.10}$. On the observational side, 
\cite{mokiem07} derived a relation for early-type stars in our Galaxy and 
in the Magellanic Clouds ($Z / Z_{\odot} = 0.5, 0.2$ for LMC and SMC, respectively), obtaining 
a relation $\dot{M} \propto Z^{0.83 \pm 0.16}$ that is consistent with the radiatively-driven wind theory. 

These conclusions have been recently questioned by \cite{tramper11}, who observed six O-type stars 
in galaxies with sub-SMC metallicities, $Z / Z_{\odot} = 1/7$ (IC 1613, WLM, and NGC 3109) with VLT/\xshooter. 
From the modeling of the optical spectrum,  they derived mass-loss rates that are higher than the values expected from the 
theoretical relation for the starsÕ metallicity. The derived mass-loss rates are more consistent 
with LMC-type metal abundances. Several uncertainties were investigated that could affect these
determinations, including the metallicity of the galaxies (measured from the O/H
ratio) and the correction for nebular emission, or the influence of stellar multiplicity on the derived mass-loss rate by dilution or wind-wind collisions.
None of the solutions considered by \cite{tramper11}, however, could explain their unexpected findings. 
These original results have been further supported by the analysis of an extended sample of 10 stars \citep{tramper14}.

If confirmed, such a breakdown of the mass-loss rate scaling with metallicity would have dramatic
consequences on the ultimate evolution of these objects, hence far more serious consequences
than a ``simple'' revision of our general physical understanding of massive star properties.
Indeed, the lower mass-loss rates that are predicted at low metallicities imply that the removal of
angular momentum during a star's evolution is substantially reduced at lower metallicities,
increasing the effects of rotation, especially on mixing \citep{maeder01}.
How to correctly treat the effects of rotation and metallicity on mass-loss and mixing
is relevant to any field that needs to understand the life cycles and feedback of massive stars.
These stars are the primary source of production of all elements heavier than O (except Fe),
that they release in the interstellar medium via their stellar winds or when they explode as
core-collapse supernovae. The chemical and dynamical evolution of galaxies is thus set by massive stars \citep{freyer03}. Another example is
the collapsar model for long-duration gamma-ray bursts \citep{woosley93}, which requires evolved massive progenitors
with large amounts of angular momentum and mixing, possibly to the point of completely mixing the star to a
homogeneous composition \citep{yoon05, martins13}.
Furthermore, the WR/O star ratio in classical
single-star evolution models is also expected to be modified by the metallicity dependence
of the winds: low-metallicity environments should produce less WR stars,
which therefore impacts the number of type Ib/c supernovae \citep{meynet05}. 
On the contrary, should stronger winds prevail at low metallicity, this would imply lower numbers of collapsars, and of type Ib and
Ic supernovae in the high-z Universe.

\cite{tramper11, tramper14} based their analysis of the mass-loss rate mostly on the H$\alpha$ line.
Several studies have shown that the sensitivity of H$\alpha$ is limited to winds with mass-loss
rates greater than about 10$^{-7}$ \msunyr, and this line only shows  core filling for mass-loss rates
between 10$^{-6}$ and 10$^{-7}$ \msunyr\ \citep[see e.g.][]{marcolino09}. Additionally, wind clumping
was neglected, although the presence of clumps has been established even at low
metallicities \citep[see][and refs therein]{puls08}. \cite{tramper11, tramper14} argue, however, that they expect that their conclusions would remain unchanged, 
should clumping be included in the analysis. 
This statement is quite surprising since, H$\alpha$ being sensitive to the square of the wind density,
neglecting clumping leads to overestimate the mass-loss rates by a factor 1/$\sqrt{f}$ (where $f$ is
the volume-filling factor associated with clumping). This corresponds to factors
of 3 to 10 that are of similar amplitude than the expected metallicity effect between a Galactic massive star and a counterpart with the exact same 
photospheric parameters in the SMC typically. 
This raises the question of the reliability of the optical diagnostic for mass-loss determination, as well as of the underlying metallicity adopted for the galaxies IC 1613, WLM and NGC 3109. 

Far-UV  (FUV) spectroscopy, on the other hand, is the adequate tool to fully address and resolve this outstanding
issue of the dependence of hot, massive star mass-loss rates with metallicity.
Spectroscopy in the 1150 -- 1800 \AA\ waveband provides many of the major wind line diagnostics, 
including the \nv\ \lb\lb1238, 1242 doublet that is also temperature, abundance, and X-ray ionization dependent, the \civ \lb\lb1548, 1550 
doublet that is particularly useful to diagnose weak winds, and the ÓBalmer alphaÓ transition of \heii\ (\lb1640) that provides an extremely
valuable clumping and abundance diagnostic for stars with strong winds \citep{hillier03}.
In addition, \niv \lb1718, \siiv \lb\lb1395,1402 (also luminosity sensitive) and \ov \lb1371, are
also sensitive to the clumpy nature of the winds of early O stars.
Many other useful diagnostics are also found in this spectral range, such as \teff\ indicators from \mbox{C}, \mbox{N}, \mbox{O}, and
\mbox{Fe} lines that can be used instead of the classical helium ionization balance \citep{heap06}.
Several lines are also good indicators of \mbox{CNO} and iron abundances \citep[e.g.][]{hillier03,
bouret03}. The numerous iron lines (\feiii\ to \fev) present in the FUV spectrum provide a direct measurement of 
the stellar metallicity. 

The goal of the work presented here was to corroborate (or not) the breakdown of the mass-loss rate -- metallicity relation at sub-SMC metallicities with 
FUV observations with HST/\cosp\ of the three brightest stars from the original study by \cite{tramper11}.
The paper is organized as follows: we present the observations in Sect. \ref{sect_obs} and we discuss the spectra morphology in Sect. \ref{sect_morpho}.
We describe the model and method we used for the determination of the stellar parameters in Sect. \ref{sect_model}; we present the results of this modeling in Sect. \ref{sect_result}.
Then, in Sect. \ref{sect_metal}, we discuss what the actual metallicities of the galaxies are, while a possible binary status for the targets is presented in Sect. \ref{sect_binary}.
Systematic uncertainties in the mass-loss determinations are discussed in Sect. \ref{sect_error},  and wind properties at low metallicity in Sect. \ref{sect_obs_wind_prop}. 
In Sect. \ref{sect_pred_wind_prop}, we discuss how the wind properties we derive compare to theoretical expectations.
We summarize our conclusions in Sect. \ref{sect_conclusion}.

\section{Observations}
\label{sect_obs}
We used the Cosmic Origins Spectrograph (\cosp) onboard the Hubble Space Telescope (HST),
to obtain high-resolution (R $\approx$ 20,000) FUV spectra of the three brightest stars from \cite{tramper11}, namely
A13 and B11 in IC 1613 and A11 in WLM \citep[see also][]{bresolin06, bresolin07}. Aside from their intrinsic brightness, other selection criteria
were that
 {\it i)} IC1613-A13 is the only star in the \cite{tramper11, tramper14} sample with a mass-loss rate (\mdot) consistent with the theoretical prediction; 
{\it ii)} A11 is one of the only two stars in \cite{tramper11} with a mass-loss rate unambiguously larger ($\sim$ an order of magnitude) than predicted, but the other
object is almost one magnitude fainter; 
{\it iii)} A11 and B11 are late-type O supergiants ($\sim$ O9.5 I) with much cooler effective temperatures (\teff\ $\approx$ 30, 000 K) than A13. Interestingly, \mdot\ for
B11 could be compatible with theory within the errors. On the other hand,  A11 is notably
brighter than B11 (0.35 dex), has lower surface gravity (0.2 dex) and a higher mass-loss rate (0.6 dex), suggesting that it is more evolved than B11. 
Observing both stars in the FUV with HST/\cosp\ would allow us to further constrain their evolutionary properties and derive unambiguous values for the mass-loss
rates and address the issue of clumping in low metallicity stars.

Our program GO 12867 (PI: T. Lanz) was granted 18 orbits. Each target was observed in a sequence of 6 science exposures
with \cosp\, using gratings G130M (\lb1291, \lb1327) and G160M (\lb1577, \lb1623).
This sequence provides coverage without gap of the FUV spectrum between 1132 and 1798 \AA.
Overlapping segments have been co-added to enhance the final signal-to-noise ratio (SNR). 
Thanks to the high efficiency of \cosp\ in the FUV and little extinction toward the host
galaxies, we achieved typically SNR $\approx$ 25 per resolution element, in 6 HST orbits for each target
(2 orbits for G130M and 4 orbits for G160M; the latest grating covers the \civ\ \lb\lb1548,
1550 doublet and \heii\ \lb1640). 
Informations about the targets, including UBVRI photometry and exposure times are summarized in Table \ref{tab0}. 

 \cosp\ spectra are flux-calibrated \citep[see][for details and limitations]{massa13} providing the flux distribution in the FUV 
wavelength range. We shall use the spectral absolute fluxes, together with UBVRI photometry (Table \ref{tab0}) to construct spectral energy distributions (SEDs) and
constrain the intrinsic stellar luminosity (see Sect. \ref{param_sect}). It is possible, however, that the spectra of the three targets are affected by physical or line-of-sight, 
unresolved multiplicity. 
Indeed, the nominal primary science aperture of \cosp\ has a radius of 1.25 arcsec\footnote{Note that the \cosp\ $\it handbook \rm$ \citep{holland12}
mentions that ``the aberrated beam entering the aperture allows objects up to 2 arcsec from the center of the aperture to contribute to the recorded spectrum''}.
At the distances of IC1613 and WLM (cf. Table \ref{tab0}), this subtends a field of view with physical size of 
9-12 pc across, several times the size of a cluster like R136 \citep[e.g.][]{crowther10}.

On the other hand, the FUV range of a spectrum is dominated by the hottest objects. Shall hotter objects than the targets be in the \cosp\ entrance aperture, the spectra
will generally present signs of contamination (e.g. lines from higher ionization stages) which is not seen (see Sect. \ref{sect_morpho}). 
More importantly, if the stars are in binary systems and their mass ratios differ greatly from unity, then the derived physical properties reflect those 
of the primary\footnote{But see  R136A for a case not showing strong evidence, from spectroscopy and the SED, for binarity even though its
a multiple system.}.

Furthermore, ground-based observations with spatial resolution indicate that the two targets in IC1613 are isolated within the \cosp\ field of view. We note also that
\cite{tramper11, tramper14} found that their luminosities are compatible with those of (single) stars with same spectral type/luminosity class in the MCs \citep[e.g.][]{massey09}. 
Therefore, we are confident 
that the physical parameters we can derive from the modeling of the FUV (+ optical) spectra truly reflects the properties of the targets. 
The situation is not as clear for WLM-A11, where a significant source of optical flux is present within a 2.5'' diameter circle (hence the ``a'' subscript of the luminosity class) in optical images. 
Since the flux level in the FUV, plus the whole UBVRI photometry, is also quite high, it might indicate that this target is actually a binary system with components
of similar spectral types (more exactly similar effective temperatures). We will come back on this point in Sect. \ref{sect_binary}.

 \begin{table*}
\centering \caption{Basic parameters of our targets and exposure times of the \cosp\ spectra.} 
\begin{tabular}{llcccccccccc}
\hline
 Star			&	Sp. Type	& $U$	&	$B$		&	$V$		&	$R$		&	$I$	& $E(B-V)$	& dist. 	& $M_{V}$	&  \multicolumn{2}{c}{Texp (s)} 		   \\ 
 			&			&		&			&			&			&		&			& (kpc)	&			& G130M	&	G160M	\\
 \hline
IC 1613-A13	& O3-O4 V((f))	&   17.66	&	18.73	&	18.96	&	...		& 19.26 	&	0.03	 	& 721	&  -5.42		&	4640.	&  10006.\\	
IC 1613-B11	& O9.5 I		&  17.45	&	18.49	&	18.62	&	...		& 18.78	&   	0.08		& 721	& -5.92		&	4581.	&  9927. \\  
WLM-A11		& O9.7 Ia		&  17.23	&	18.27	&	18.38	&	18.45 	& 18.57	&	0.04		& 995	& -6.73		&	4625. 	&  9992. \\	
  \hline
\end{tabular}
 \label{tab0}
    \begin{list}{}{}
\item[Note :]    Spectral types are from \cite{tramper11} and \cite{garcia14}, while the photometry is taken from \cite{garcia09, garcia14} for the stars in IC1613 and from \cite{massey07} for WLM-A11. Color excess $E(B-V)$ is calculated using the calibration by \cite{martins06} for Galactic Stars.
Absolute magnitude are calculated using the quoted distances to IC1613 and WLM from \cite{Pietrzynski06} and \cite{Urbaneja08} respectively. 
\end{list}
   \end{table*}
 
We also retrieved the optical, VLT/\xshooter\ spectra of the three targets \citep[first presented in][]{tramper11} from the Phase 3 ESO  archives. 
These spectra have been processed using version v2.2.0 of the \xshooter\ pipeline. 
We used the spectra obtained in the UVB (300--550 nm) and VIS (550--1020 nm) arms with slit width of 0.8'' and 0.9'', which provide 
a spectral resolving power $R = 6200$ and $R = 7400$, respectively. Typical SNR per resolution element in the UVB spectra are 25, 20 and 18 
for IC 1613-A13, IC 1613-B11 and WLM-A11, respectively. In the VIS arm, SNR per resolution element ranges from 10 to 12.

Following \cite{tramper14}, we corrected the spectra from the nebular emission in the Balmer lines (and \hei\ for IC 1613-A13), with special
care for the H$\alpha$ line, as this is the principal mass-loss diagnostic available in the optical.
These spectra have been used in the modeling process to better constrain the photospheric parameters (especially \teff\ and \logg) and to check how
the synthetic H$\alpha$ profiles produced by FUV-based \mdot\ compare to observations. 


\section{Spectra Morphology} 
\label{sect_morpho}
In a first step, we compared the \cosp\ spectra of IC 1613-A13, IC 1613-B11 and WLM-A11 with existing FUV spectra 
of SMC, LMC and Galactic stars of similar spectral type and luminosity class. In principle, 
this  should provide a direct, qualitative, model-independent, indication about the mass-loss rate -- metallicity relation. 
The line of reasoning is straightforward. The  strength of a P~Cygni profile is directly related to the optical depth of the wind in the corresponding line. 
The optical depth is related to the total number of absorbing ions  in the transition giving rise to the spectral line and can be written as
$\tau_{rad} \propto \dot{M}q_{\rm i}A_{\rm E}$ with $A_{\rm E}$ and $q_{\rm i}$ the abundance of element E and its ionization fraction for stage i.  It is therefore
readily seen that, for stars with similar physical properties and evolutionary stages (as should be for stars within the same spectral/luminosity classes), hence yielding similar 
ionization fractions, $\tau_{rad}$ scales with the mass-loss rate and the abundance (which is expected to scale with metallicity).  

\begin{figure*}
\centering
\includegraphics[scale=0.41, angle=0]{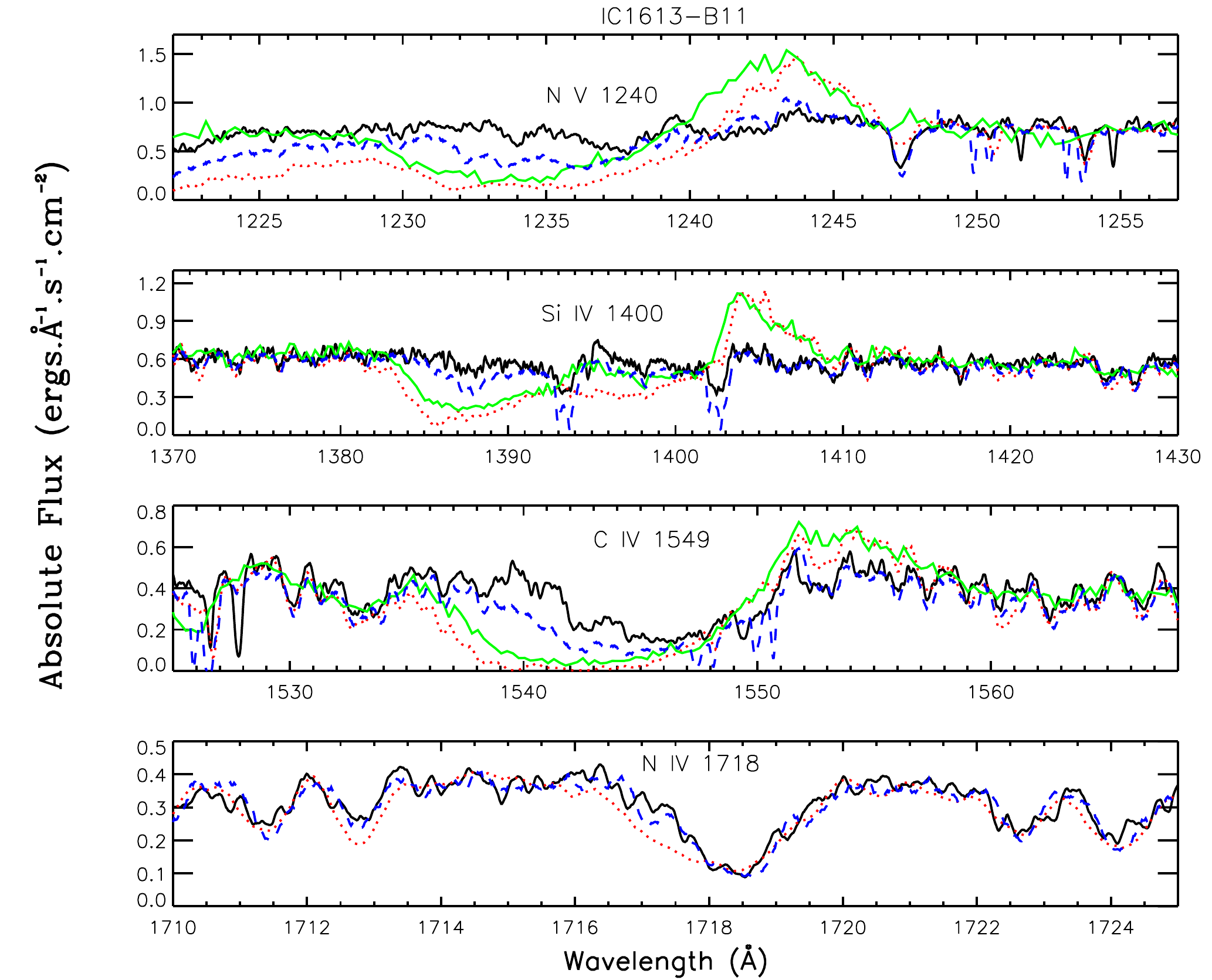}
\includegraphics[scale=0.41, angle=0]{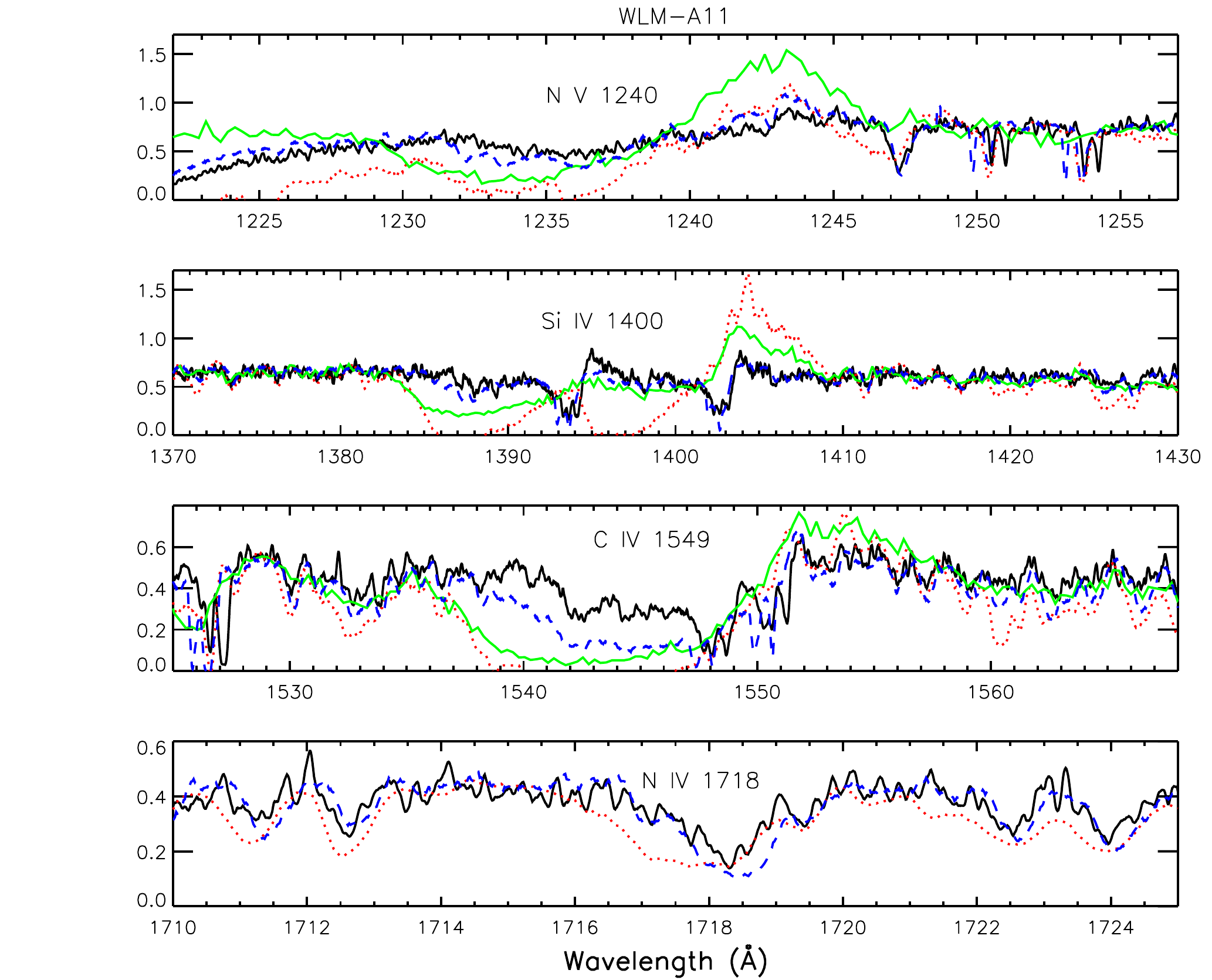}
\caption[]{P~Cygni profiles for B11 (left) and A11 (right) in black. Overplotted in red, green and blue are the spectra of the Galactic, LMC and SMC stars 
(respectively) chosen for comparison  (cf. Table \ref{tab1}). For plotting purpose, the fluxes were scaled by a factor 10$^{14}$.
}
\label{fig1}
\end{figure*}
We selected  stars with spectral classification as close as possible to those of  IC 1613-A13, IC 1613-B11 and WLM-A11 although a perfect match was not
always possible. Also, whenever possible, we chose stars with high resolution spectra, such that we can further compare the numerous lines from iron ions. 
High-resolution \iue\ spectra exist for Galactic massive stars of all spectral type and luminosity class. In the Magellanic Clouds (MCs), several \iue\ spectra are available but they have poor signal-to-noise ratios and were not considered for this comparison.  
In the MCs, good signal-to-noise high-resolution spectra have been obtained with HST but for a limited number of objects. 
For early-type dwarfs in the LMC, such good signal-to-noise high-resolution (Goddard High-Resolution Spectrograph, \ghrs) spectra exist but they are restricted to stars in R136. These O-type stars often seem quite extreme 
objects in terms of stellar and wind parameters \citep{dekoter98, prinja98b, massey04}, in excess of those found for Galactic stars of similar spectral type \citep{bouret05, martins12a}. 
We chose to use the spectrum of a field star instead, namely LH81W28-5, whose spectrum was obtained at low resolution though (G140L) with \stis\ \citep{massey04}. 
We could not find a high-resolution spectrum ( \cosp, or \stis\ or \ghrs) for LMC late-type supergiant and we had to rely on the more modest resolution provided by \fos\ for BI170; we use it for comparison to both IC1613-B11 
and WLM-A11. We could find only one late-type supergiant in the SMC, namely AzV 327, with high-resolution (\stis) FUV spectrum. Here again, this star is used for comparison to B11 and A11.
The IDs of the stars we used for the morphology comparisons are summarized in Table \ref{tab1}, together with additional indications about their FUV spectra, V-magnitude and reddening.

\begin{table*}
\caption{IDs, spectra and photometry for the comparison stars.}
\begin{tabular}{llclccc}
\hline
\hline
Star			& Comparison Stars - Sp. Type	& Galaxy  &  FUV spectrum   	&     V  	&   B  	&    E(B-V)$^{\mathrm{a}}$	        \\  \hline
IC1613-A13	&						& IC1613	& \cosp\ GO 12867		& 		&		&					\\ 
			& HD 93205	- O3.5 V((f))	& MW	& \iue\ SWP07959	& 7.76	& 7.76	&	0.28					     		\\
			&LH81W28-5 	- O4 V((f+))	& LMC	& \stis\ GO 8633	&13.92	&13.74	&	0.15					 		                    \\ 
			&MPG 324	- O4 V((f)) 		& SMC	& \stis\ GO 7437	&14.07	&13.79	&	0.07				   \\ 
IC1613-B11   	&						& IC1613	& \cosp\ GO 12867		& 		&		&					\\          
			& HD 76968	- O9.5 Ib		& MW	& \iue\ SWP20621	& 7.09	&7.21	&	0.38					 	                   \\ 
			& BI170		- O9.7 I		& LMC 	& \fos\  GO 5444	&13.06	&13.26	&	0.07					 			    \\ 
			& AzV 327	- O9.5II-Ibw	& SMC	& \stis\  GO 7437	&13.25	&13.47	&	0.04								                    \\ 
WLM-A11		&						& WLM	& \cosp\ GO 12867		& 		&		&					\\ 		
			& HD 152424	- O9.7 Ia		& MW	& \iue\ SWP15021	&6.31	& 6.71	&	0.66						                   \\ 
			& BI170		- O9.7 I		& LMC 	& \fos\  GO 5444	&13.06	&13.26	&	0.07					 			    \\ 
			& AzV 327	- O9.5II-Ibw	& SMC	& \stis\  GO 7437	&13.25	&13.47	&	0.04		\\	
  \hline
\end{tabular}
\label{tab1}
\begin{list}{}{}
\item[$^{\mathrm{a}}$] Calculated from B-V (this table) and (B-V)$_{0}$ from \citep[]{martins06}.
\end{list}
 \end{table*}
 
\begin{figure}
\centering
\includegraphics[scale=0.41, angle=0]{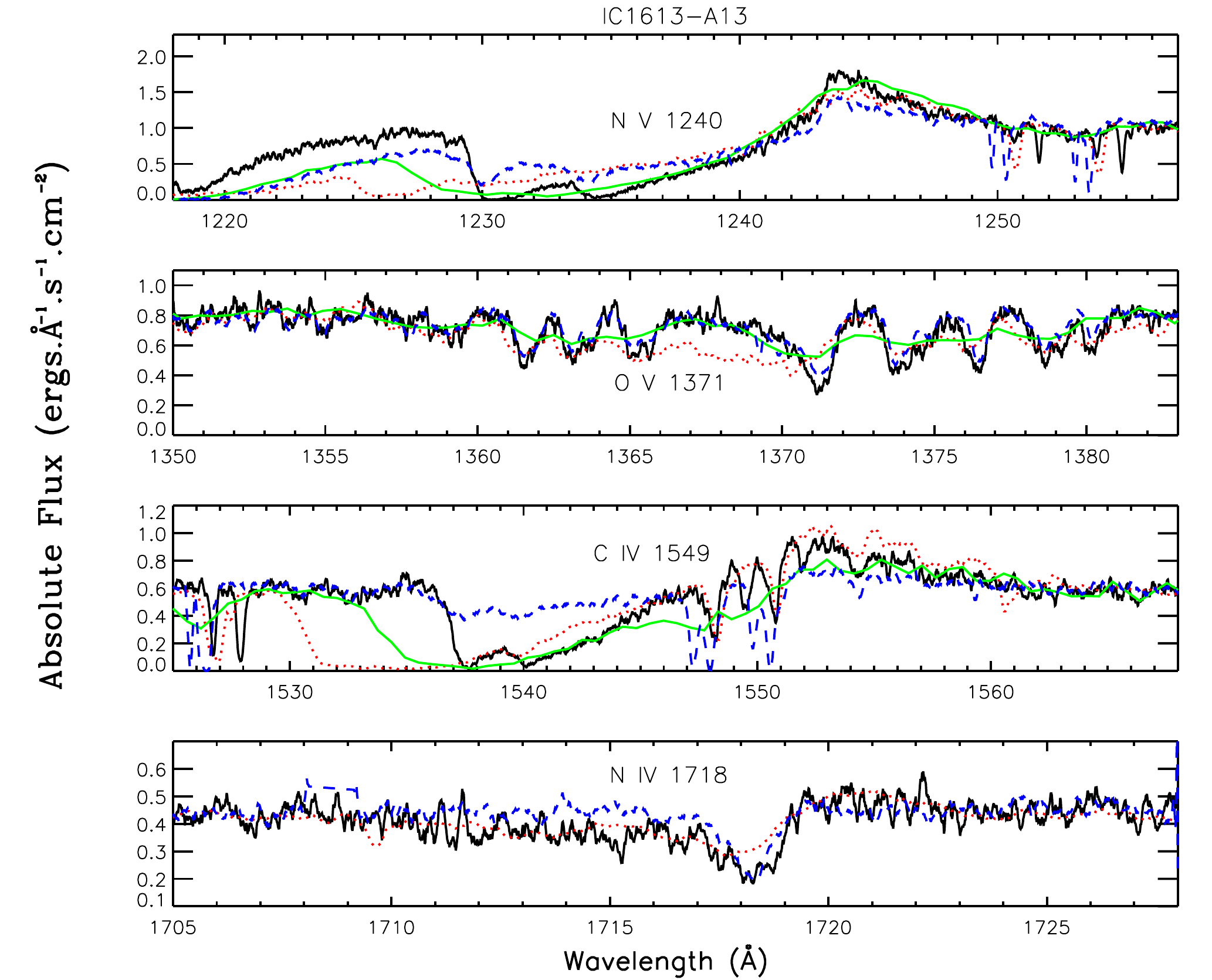}
\caption[]{P~Cygni profiles of A13 (in black). Overplotted in red, green and blue are the spectra of the Galactic, LMC and SMC stars (respectively)
chosen for comparison (cf. Table \ref{tab1}).  For plotting purpose, the fluxes were scaled by a factor 10$^{14}$.}
\label{fig2}
\end{figure}

For the comparisons presented in Fig. \ref{fig1} (B11 and A11) and Fig. \ref{fig2} (A13), we used absolute fluxes, corrected for reddening, such that we avoid potential problems 
related to the choice of a continuum to rectify the spectra. Fluxes for Galactic, LMC and SMC stars have been scaled by the appropriate ratio to match
the continua of the three targets in spectral regions of interest (i.e., around wind profiles). These scaling factors present relatively small deviation from those that would be obtained from
a scaling to the distance modulus of IC 1613 and WLM as given in \cite{tramper11}.

We first discuss the cases of IC 1613-B11 and WLM-A11 as they have similar spectral type/luminosity classes. 
Qualitatively, their behavior is easy to interpret in the framework of the line-driven wind theory. 
The top three panels in Fig. \ref{fig1} show that the wind terminal velocity decreases with decreasing metal content, from Galactic to IC 1613/WLM, while the P~Cygni profiles get simultaneously weaker.
This especially clear for the second and third panels for each star, which demonstrate that this trend holds for the resonance doublets of \siiv\ and \civ. 
On the other hand, caution should be used concerning the \nv\ resonance doublet at 1238, 1242 \AA, which is notoriously sensitive to X-ray irradiation, that is not known for the stars used for this comparison. P~Cygni profiles are well-defined for the Galactic and LMC
stars but only a weak (if any) emission component is observed for the SMC star, as well as IC1613-B11 and WLM-A11. The \civ\ \lb\lb1548, 1550 profiles still show a blueward extension that is related to
the wind whereas it is barely seen on the \siiv\ doublet because of the ten times lower abundance of Silicon versus Carbon. 
The panels showing \niv\ \lb1718 on the other hand, show very similar profiles for IC1613-B11 and WLM-A11 and their SMC counterpart, both weaker nevertheless than for the 
Galactic star of comparison. 
Overall, the morphology of the profiles from Fig. \ref{fig1} seem to contradict the conclusion by \cite{tramper11} that IC1613-B11 and WLM-A11 have mass-loss rates higher than 
expected for their metallicity. This of course assumes that the line-driven wind theory holds for stars in the Galaxy and down to SMC metallicity, as checked by \cite{mokiem07}. 

It is also easily seen on Fig. \ref{fig1} that the iron lines present in the FUV spectra of IC 1613-B11 and WLM-A11 (see e.g. the bottom two panels, \feiv\ lines between \lb\lb1530--1570 and 
\lb\lb 1710--1730)
are moderately weaker, if at all, than those of the SMC star we chose
for comparison, despite a slightly higher projected rotational velocity (\vsini) for AzV 327 (100 \kms, Bouret et al. in preparation).  Although it could be 
an indication that the iron abundance in our targets is slightly smaller than the fiducial 0.2 \fe$_{\odot}$ of SMC O-type stars, it could as well be related 
to a lower micro-turbulent velocity in their photospheres. Settling this issue requires quantitative modeling, which we will discuss later in this paper. 

Let us now discuss IC1613-A13. As already said the \nv\ resonance doublet (top panel of Fig. \ref{fig2}) is sensitive to the X-ray flux (not known for most of the stars here); 
it also often heavily blends with interstellar Ly$\alpha$ on the blue side of the profile, which is an additional issue for early-type O dwarfs as their wind terminal velocities 
are expectedly higher than for the late-type supergiants. Despite this, \vinf\ does appear to decrease with metallicity, a trend that is further confirmed on the other panels. 
Using the \civ\ \lb\lb1548, 1550 resonance doublet (third panel,), we see that \vinf\ for the LMC star ($\approx$ 2700 \kms), intermediate between \vinf\ for the Galactic 
and SMC star. Remarkably, \vinf\ for IC1613-A13 and the counterparts in the SMC (MPG 324) are very similar, which might indicate that the metallicities in both galaxies are alike. 
 
The conclusions are not as obvious for the mass-loss rate. For instance, th \nv\ \lb1240 P~Cygni profile in A13 is clearly as strong (both the absorption and emission
components) as the profile in the Galactic star (HD 93205) and in any case stronger than in MPG 324 in the SMC.
The second panel, showing \ov\ \lb1371 seems to indicate that, the LMC star left aside, we do observe a trend for
lower metallicity stars to have lower \mdot. The \ov\ \lb1371 profile for IC1613-A13 is indeed almost completely photospheric while the profile for MPG 324 (SMC) and HD 93250 (MW) 
present a blueward extension (stronger for the Galactic star). 
However, \ov\ \lb1371 is one of the FUV line that is expected to be most sensitive to wind clumping \citep{bouret05}, together with \niv\ \lb1718. The later is shown in the bottom panel and the same comments hold for both lines (this wavelength range is not covered by the \stis\ spectrum of the LMC star).
Care should be used when interpreting these profiles as their strengths also reflect the degree of wind clumping, that most likely varies from star to star. 
The gap in metallicity between our Galaxy and the SMC is large enough however, such that the weakening of the wind profile caused by 
the factor of five in metallicity is visible in the profiles in this panel. Here again, the resolution of the LMC spectrum is too low to reach a firm conclusion about the strength
of the \ov\ \lb1371 profile relative to the SMC counterpart and IC1613-A13 accordingly.

The third panel, showing the \civ\ \lb\lb1548, 1550 resonance doublet, further adds to the confusion. 
The expected qualitative trend is observed for the Galactic, LMC and SMC stars for the mass-loss rate (and terminal velocity). 
The absorption components of the P~Cygni profiles for the three stars go from clearly saturated (Galactic, then LMC) to unsaturated (SMC); the emission decreases in intensity as well. 
This is fully consistent with theoretical expectations from the radiatively-driven winds theory, but the behavior of IC1613-A13 is at odds with this predicted trend. 
Although its \vinf\ seems quite similar to that of  MPG 324 in the SMC, the intensity of its P~Cygni profile is much stronger, suggesting that the mass-loss rate is higher. 
One should remember, however, that the intensity of the P~Cygni profile is determined by
\mdot$q_{\civ}$$A_{C}$ where $q_{\civ}$ is the ionization fraction of \civ\ and $A_{C}$ the abundance of Carbon. Therefore, we cannot rule out that either the Carbon abundance or
the C$^{3+}$ ionization fraction, or a combination of both, is really lower in the SMC star such that the P~Cygni profile is weaker. 
It is also noteworthy that the \civ\ P~Cygni profile of IC1613-A13 shows well-defined narrow absorption components (NACs) that are 
useful to measure the wind terminal velocity \citep{prinja98b}. DACs are also observed in several SMC O dwarfs  of early spectral type \citep{bouret13}, where the absorption component is never saturated (see Fig. \ref{fig2}, the case of MPG 324) .


To summarize this section, we find that the P~Cygni profiles of  IC 1613-B11 and WLM-A11 are compatible with the winds of these stars being weaker than winds
of stars with similar spectral types and higher metallicity. Since \cite{mokiem07} showed that the wind quantities of massive stars in the Milky Way, the LMC and SMC follow a trend 
in quantitative agreement with the predictions of the radiatively-driven wind theory, it is tempting to conclude that the spectra of our targets also comply quantitatively 
with the theoretical expectations. 
The case of IC 1613-A13 is different and only ambiguous conclusions about the wind quantities and their dependance to metallicity
is obtained from a qualitative comparison. 
We will now discuss the quantitative aspects of this conclusion, from the modeling perspective.

\section{Models and method}
\label{sect_model}
For a more quantitative view of the wind properties, we  compared the observed FUV + optical spectra of IC 1613-A13, IC 1613-B11 and WLM-A11 to synthetic spectra calculated with
the NLTE stellar atmospheres code \cmfgen\ \citep{hillier98}. 
\cmfgen\ computes NLTE line-blanketed model atmospheres, solving the radiative transfer and statistical equilibrium equations in the comoving
frame of the fluid in a spherically-symmetric outflow. The models we computed account for the presence of the following species: \mbox{H, He, C, N, O, Ne, Mg, Si, P, S, Cl, Ar, Ca, Fe, Ni}, 
representing more than 7500 full levels (more than 2000 super-levels) and $\approx$ 135,000 lines. The density structure is described as an hydrostatic part, in the quasi-static photospheric layers, smoothly connected to a $\beta$ velocity law in the wind. The radiative acceleration is calculated from the solution of the level populations, and is used to compute a new inner structure (connected to the same  $\beta$ velocity law). The mass-loss rate, density, and velocity are related via the continuity equation. After convergence, a formal solution of the radiative transfer equation was computed in the observerÕs frame \citep{busche05}, thus providing the synthetic spectrum for a comparison to observations. 

We started from the stellar and wind parameters provided in \cite{tramper11} and tuned them as needed to get 
the best fit of the FUV and optical spectra together. The influence of clumping on the wind profiles was considered. 
Clumping is described in the framework of the optically-thin clumps formalism implemented in \cmfgen. The clumped wind density, $\rho(r)$, is related to the  homogeneous (unclumped) wind density 
$\bar{\rho}$, and the volume filling factor $f$ by  $\rho(r)=\bar{\rho}/f$. The filling factor decreases exponentially with increasing radius (or, equivalently, with increasing velocity):
$f = f_\infty + (1-f_\infty) \exp(-v/v_{\rm cl})$
where $v_{\rm cl}$ is the velocity at which clumping starts. More physical description of clumping exist nowadays and will be discussed in Sect. \ref{sect_clumping}.
A  limitation of this parameterization is that {\it f} is monotonic whereas theoretical simulations by \cite{runacres02} and observational studies \citep{puls06, najarro11} suggest it is non-monotonic. 
We found no need to tune $v_{\rm cl}$ to improve the fit to observed lines, hence we adopted a canonical $v_{\rm cl}$ = 30 \kms\ \citep{bouret05}.  

The major consequence of this
formalism is that mass-loss rates derived with clumped models are reduced by a factor $\sqrt(\finf)$, where $\finf$ is the volume filling-factor associated with clumping, 
compared to mass-loss rates obtained from homogeneous models \citep[e.g.][]{bouret05}. In the rest of the paper, we will refer to the ``clumped'' mass-loss rate
(\mdot\ and \finf\ derived from the modeling of the FUV spectra) to compare to theoretical quantities and/or mass-loss rates from \cite{tramper14}. 

We also accounted for the effects of shock-generated X-ray emission in the model atmospheres. Since no measurements are available for IC 1613-A13, IC 1613-B11 and WLM-A11 in the
X-ray wavelength range, we adopted the standard Galactic luminosity ratios log$\rm L_{X}/L_{bol} \approx -7$ \citep{sana06}, 
with the bolometric luminosity determined in the present study. We acknowledge however that there is no guarantee that this scaling ratio should prevail at low metallicity.  
\cite{owocki13} suggested for instance that this ratio should change with the physical nature of the shock generating the X-rays (radiative or adiabatic) and the wind density. 
Depending on the actual relation between the metallicity and the mass-loss rate,  log$\rm L_{X}/L_{bol}$ could change accordingly. 

\subsection{Stellar and wind parameters}
\label{param_sect}
The photospheric (including surface abundances) and wind parameters have been derived following the same methodology as in previous papers \citep[see e.g.][]{bouret12, bouret13}. 


\begin{itemize}
\item {\bf Stellar luminosity}:
The distance to IC 1613 and WLM were taken from \cite{Pietrzynski06} and \cite{Urbaneja08}, respectively. We used the flux-calibrated \cosp\ spectrum, providing the flux distribution in the FUV 
wavelength range, together with optical photometry (cf. Table \ref{tab1}) to build spectral energy distribution (SED). The intrinsic stellar luminosity 
was then derived, by fitting the SED. Note that we used the color excess quoted in Table \ref{tab1} to take reddening into account, using the data by \cite{cardelli89}. 

\item {\bf Effective temperature and surface gravities}:
\teff\ was constrained from the helium (primarily \hei\ \lb4471 to \heii\ \lb4542 equivalent width ratio, as well as by fitting the line profiles) in the optical spectra.
We add that the relative strength of helium lines was also used to constrain the helium abundance.  
 Ratios of lines in the FUV spectra \citep[e.g.]{heap06, bouret13} were also used for further consistency, e.g. \civ\ \lb1169 to \ciii\ \lb1176, 
 or \niii\ \lb\lb1183-1185 and \niii\ \lb\lb1748-1752 to \niv\ \lb1718 ratios. This is particularly important for the hottest star of the sample, namely A13, where the \hei\ lines become weak 
 \citep[see also][for the use of nitrogen optical lines to constrain \teff\ for the earliest O-stars]{rivero12}.
The \civ\ \lb1169  to \ciii\ \lb1176 ratio is especially useful in early O-type stars because of its very good sensitivity to \teff\ and weak dependance on gravity; hence we used it as the primary \teff\ diagnostic in the UV.

The Stark-broadened wings of Balmer lines, especially H$\gamma$, were the primary constraint on \logg. 
The S/N of the \xshooter\ spectra is such that it is not possible to achieve a typical accuracy (1 $\sigma$) better about 0.2~dex in \logg. 

\item {\bf Microturbulence}:
A constant microturbulent velocity of 15 \kms\ was adopted for calculating the atmospheric structure. 
To calculate the emerging spectrum, we used a radially dependent turbulent velocity \citep[see also][]{hillier98}. 
The microturbulence in the photosphere \xit\ was set mostly using the \sv\ \lb1502 line, that is very sensitive to this parameter over a large range of \teff. We further tuned it if necessary using the known 
sensitivity of some \he\ lines to \xit\ \citep{villamariz00, hillier03}. 
In the outer region of the wind, the turbulence was adopted to fit the shape and slope of the blue side of the absorption component of the \civ\ 1548-1551 P~Cygni profile. 

\item {\bf Surface abundance} :
The thousands of photospheric lines of iron ions (\feiii\ to \fevi) present in the FUV are most useful to constrain the metal content of the stars. 
This shall provide an independent check of the stellar metallicity so far inferred  from the ISM O/H ratio in the host galaxies of our targets. 
While the lines of  \mbox{Fe} ions grow stronger with increased gravity and higher iron abundance and/or microturbulent velocity, the ratio of line strengths is
most sensitive to temperature \citep[see the discussion in][]{heap06}. Practically, we used \teff, \logg, and \xit\ determined as we just presented to
derive an iron abundance as reliable as possible. Abundances of  $\alpha$-elements included in the models (\mbox{Ne, Mg, Si, S, Ni}) were scaled down from the solar 
values of \cite{asplund05}, adopting the same factor as for iron.
Note that for the specific case of \si, we tuned the silicon abundance in the two late-type O supergiants, using several lines of 
\siiii\ (triplet at \lb\lb\lb1294.54, 1298,89, 1298.96 or the singlets at \lb1312.59 and \lb1417.24) present  in the \cosp\ spectra.

-- {\it Carbon abundance}:
\civ \lb1169, \ciii \lb1176 are prime indicators for carbon abundances. These lines proved to be sensitive 
to \teff\ and \xit\ and \ciii \lb1176 further showed sensitivity to the wind density for the two stars with later spectral types. In this case, we gave more weight
to  \ciii \lb1247, which is clearly detected in IC1613-B11 and WLM-A11 and not affected by the adjacent, weak, \nv\ resonance doublet. 

The VLT/\xshooter\ spectra turned out to be quite noisy and with a spectral resolution somewhat too low, such that \ciii\ \lb\lb4068, 70 were used only as secondary
indicators for C/H determination for the late-type supergiants (B11 and A11). Besides, we did not include
\ciii\ \lb\lb\lb4647, 50, 51 although is clearly detected in the spectra of B11 and A11. \cite{martins12b} indeed showed that it is
very sensitive to the adopted atomic physics and other modeling assumptions\footnote{We note however that the influence of \fe\ lines found by \cite{martins12b} should be less than for galactic O stars because the iron (and more generally metal) abundance is expected to be down by a factor of five or more}.  

-- {\it Nitrogen abundance}: 
In the UV, the photospheric lines \niii\ \lb\lb1183-1185 and \niii\ \lb\lb1748-1752  are used as primary diagnostics.  
\niv\ \lb1718 is also used for the late supergiants, but not for A13, as it is clearly affected by the stellar wind in this early-type star. 
In the optical \niii\ \lb\lb4634-42 and \niv\ \lb4058 are the strongest nitrogen lines in the spectrum of A13, and were used as additional diagnostics for nitrogen abundance.
Their formation processes, including their sensitivity to the background metallicity, nitrogen abundance, and wind-strength, have been discussed extensively by \cite{rivero11, rivero12}.

-- {\it Oxygen abundance}:
The \oiv\lb\lb1338-1343 lines usually serve as a diagnostic for O/H in the UV. Here however, they show a blue asymmetry that indicates a contribution from the wind for the earliest star (A13) with 
the stronger wind, and they are very weak (if detected at all) in the late-type supergiants. \oiii \lb\lb1150, 1154 could be used instead but these
lines are on the shortest side of the UV spectra and the flux level there is more uncertain. Similarly, \ov\ \lb1371 is detected for the hottest star (A13) but it
is very sensitive to the to the mass-loss rate and clumping parameters. 
Because the number of useful/reliable oxygen lines is so limited,  the oxygen abundance was first set such that  the sum of \mbox{CNO} is conserved, with respect 
to an initial mixture \mbox{CNO} scaled to the metallicity of the star. It was then tuned if necessary to improve the fit quality to \oiv \lb\lb1338, 1343.


\item {\bf Wind parameters}:
The wind terminal velocities, \vinf, were estimated from the blueward extension of the absorption component of FUV P~Cygni profiles.
Although the P~Cygni profiles of the three targets are not saturated, narrow absorption components are visible, especially in \nv\ and \civ\ profiles and were used to derive estimates of \vinf, following
the approach discussed in \cite{prinja98b}.
The typical uncertainty in our determination of \vinf\ is 100 \kms\ (depending on the maximum microturbulent velocity we adopted).

Mass-loss rates were derived from the analysis of UV P~Cygni profiles, such as \civ\ \lb\lb1548, 1551, \ov\ \lb1371
and \niv\ \lb1718 for A13 and \siiv\ \lb\lb1394, 1403 and  \civ\ \lb\lb1548, 1551 for B11 and A11, respectively.
We emphasize again the strong dependance of the \nv\ resonance doublet on the adopted X-ray luminosity, especially for the stars
with cooler \teff. As log$\rm L_{X}/L_{bol}$ is not known for our targets, we did not use \nv\ \lb1238, 1242 P~Cygni profiles to constrain \mdot.

The $\beta$ exponent of the wind velocity law was derived from fitting the shape of the P~Cygni profile. 
The clumping parameters, $\finf$ and $v_{\rm cl}$ were derived in the UV domain. We used \ov\ \lb1371 and \niv\ \lb1718 \citep{bouret05}, as well as
\siiv\ \lb\lb1394, 1403 and  \civ\ \lb\lb1548, 155. All these lines indeed present significant sensitivity to clumping in the temperature regime of our targets.
Some photospheric lines in the optical presented some sensitivity to the adopted filling factor (and scaled \mdot). For photospheric \hi\ and \he\ lines,
for instance, this is essentially caused by a wind contribution that is weaker in clumped models, which produces deeper absorption than smooth-wind models.
   
\end{itemize} 
Synthetic spectra have been convolved with gaussian instrumental profiles, at the resolution delivered by HST/\cosp\ and VLT/\xshooter, respectively.  
Then, we further convolved these spectra with a rotational profile to account for the projected rotational velocity of the star. 
For simplicity, we adopted the values listed \cite{tramper11} and checked on several FUV line profile for consistency \citep[see e.g.][]{bouret13}. 
Given the moderate spectral resolving power and signal-to-noise ratio of the VLT/\xshooter\ spectra, we did not try to derive additional 
broadening mechanisms using the classical Fourier Transform method \citep{simon10}. 

\section{Results of the modeling}
\label{sect_result}
The best-fit of the FUV and optical spectra of the targets are presented in the Appendix, while the parameters of the corresponding model are listed in Table \ref{tab2}.  
We emphasize that the values we quote for \mdot\ in this table were obtained by fitting the FUV spectra we presented hereabove, but they do provide consistent fits
to the optical spectra. For comparison to parameters derived from optical spectrum, we used the most recent work by \cite{tramper14} as the fitting procedure 
presented there was updated compared to \cite{tramper11} to account for the scaling of \vinf\ with metallicity, and the qualitative conclusions already reached 
in \cite{tramper11} are unchanged.

\subsection{IC 1613-A13}
\label{sect_res_a13}
We find A13 to be cooler than in \cite{tramper14}, although both \teff\ overlap within the respective error bars. 
Since we used the ionization balance of \civ\ and \ciii\ as well as those of the iron ions present in the FUV to constrain the parameters, our \teff\ estimate is therefore less prone to uncertainties 
related to the weakness of \hei\ lines in the optical for such an early-type O star. We note that our value is very close to \teff\ derived by \cite{garcia13} from the modeling with \cmfgen\ of a 
lower resolution \cosp\ spectrum of this star.  
The luminosity we derive, based on SED fitting, is a mere 0.09 dex smaller than in \cite{tramper14}. On the other hand, we favor a surface gravity \logg\ = 3.75 over the lower value 
\citep[\logg\ = 3.65 in][]{tramper14}, although the error bars are quite large in both case. This higher value stems mostly from the influence of \logg\ on the global ionization balance
\citep[see][]{heap06} and is also more compatible with the spectral/luminosity class of IC 1613-A13. As a consequence, the stellar, spectroscopic, mass we derive is in good agreement 
with the mass quoted by \cite{tramper14}. 

The helium fraction (by number) we derive is a factor of 2.5 times smaller than in \cite{tramper14}.  We find that such a high helium fraction as theirs would produce 
\heii\ \lb1640 with a well-developed P~Cygni profile, as well as strong wings in emission for the \heii\ \lb4686 line (see Fig. \ref{fig8}). 
None of those are observed and we discarded this solution; it is also not consistent with the evolutionary status of this star. 

\begin{figure}
\centering
\includegraphics[scale=0.5, angle=0]{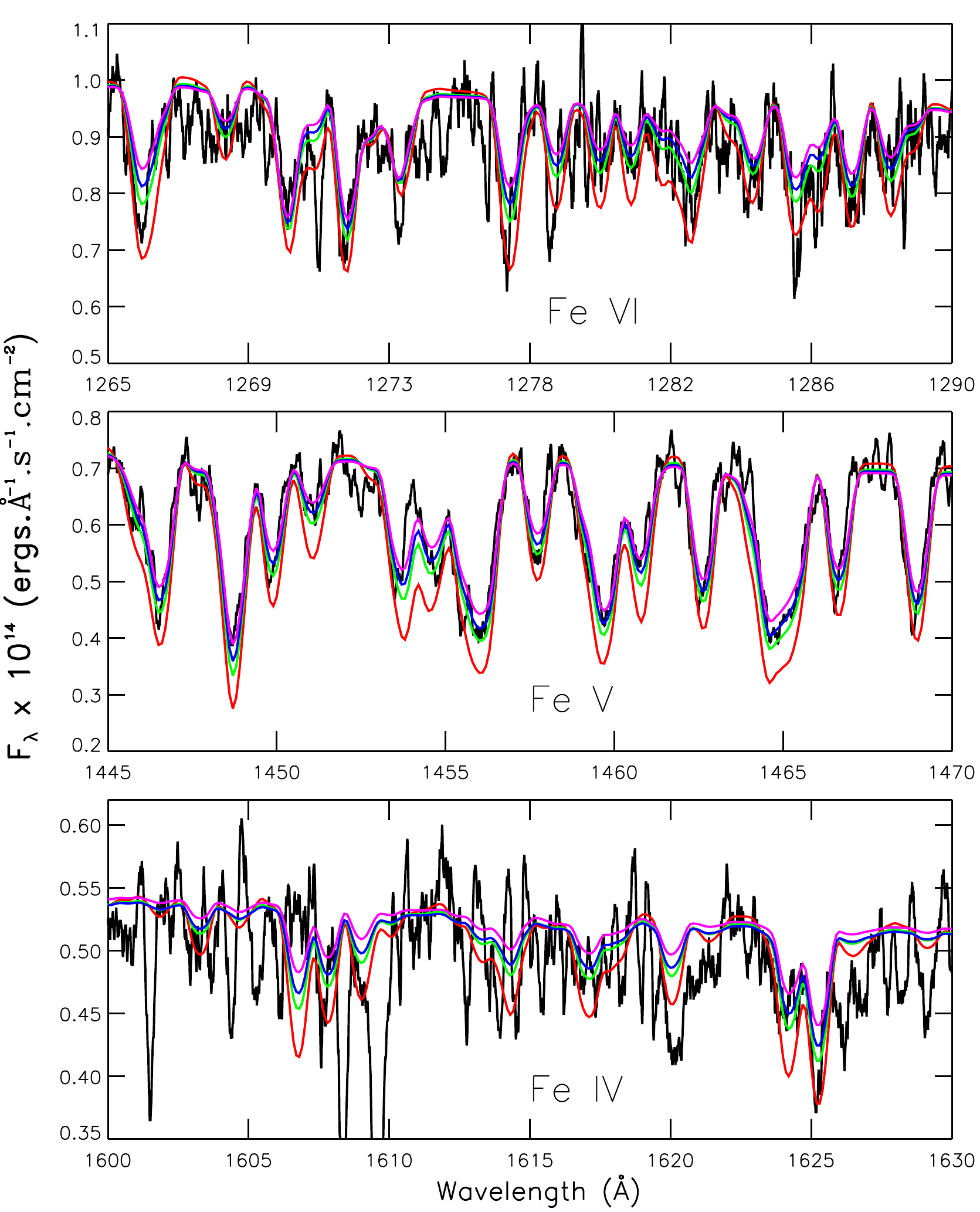}
\caption[]{In black, line profiles for different transitions of iron ions (\fevi to \feiv\ from top to bottom) in the FUV range, observed for IC1613-A13.  Overplotted in red, 
green, blue and pink are profiles obtained for models with the same parameters but the iron abundance, ranging to 1/2, 1/5, 1/7 and 1/10 $\fe_{\odot}$ respectively.}
\label{fig3}
\end{figure}

The iron abundance, which we shall now refer to as the global metallicity,  was first set at $\fe/\fe_{\odot}=0.14$ \citep[][and refs. therein]{tramper11}. 
Models with an iron content typical of SMC ($\fe/\fe_{\odot}=0.20$) and LMC ($\fe/\fe_{\odot}=0.50$), as well as $\fe/\fe_{\odot}=0.10$, were computed. The higher and lower metallicity options are easily discarded 
but on the other hand, it is not possible to rule out a SMC-like metallicity for IC 1613-A13. As already outlined in Sect. \ref{sect_model}, the strength of the many \feiii\ to \fevi\ lines 
(as well as \sv\ \lb1502) depends on the adopted abundance as much as on the photospheric parameters \teff, \logg\ and the microturbulence \xit\ set in the spectrum calculation (cf. 
Fig. \ref{fig3}). 
Within error bars of $\pm 5$ \kms\ for \xit, these lines are well fitted either adopting $\fe/\fe_{\odot} = 1/5$ or $\fe/\fe_{\odot} = 1/7$ for the iron content. The same conclusion holds for
the \siiv\ resonance doublet and the \sv\ \lb1502 line, which are purely photospheric in IC1613-A13.

An important conclusion of our study is that   IC1613-A13 is carbon/oxygen-depleted and nitrogen-rich, compared to scaled-solar \mbox{CNO} abundances.
The carbon abundance is very similar to the one we derived for MG 324, the SMC O4 V star we used for comparison in Fig. \ref{fig2} \citep[see also][]{bouret13}, but 
the nitrogen abundance is higher. The rotation rate of A13 is only mildly higher than in MPG 324 (we adopted \vsini\ = 98 \kms\ for A13 as in Tramper et al. 2011, while \vsini\ = 70 \kms\ for 
MPG 324). Furthermore, we note that IC 1613-A13 is more luminous and has lower surface gravity than MPG 324 (and other O4 V stars in the SMC for that matter), 
which points to a more advanced evolutionary status. An interpretation of the different abundances of these stars would require comparisons to predictions from
evolutionary models, which is beyond the scope of this paper. 
In any case, the similar carbon abundance of IC 1613-A13 and MPG 324 implies that the observed difference in the \civ\ resonance profiles of these stars truly reflects a difference in mass-loss rates and wind clumping (\teff\ is roughly the same for the two stars, hence the ionization fraction of \civ). The oxygen abundance conserving the sum of \mbox{CNO}, with respect to an initial \mbox{CNO} mixture that has the same scaling from the solar \mbox{CNO} mixture than the iron content of the model, produces a good fit to the FUV oxygen lines (\oiv\ \lb\lb1338, 1342 and \ov\ \lb1371) and was adopted for the rest of this work.  
 
 \begin{table}
\caption{Stellar parameters derived from the spectroscopic analysis}
\begin{tabular}{lccc}
\hline
\hline
Properties				& A13			& B11			& A11	\\ \hline
Spectral type			& O3 V((f)			& O9.5 I			& O9.7 Ia \\
\teff\ (K)				& 42500.			& 30000.			& 29000. \\
\logg\ (cgs)			& 3.75			& 3.25			& 3.25 \\
$\log \frac{L}{L_{\odot}}$	& 5.62			& 5.45			& 5.69 \\
\msp\  [\msun]			& 27.6$\pm13.$	& 22.3$\pm7.$		& 53.2$\pm9.$ \\
\xit\ [\kms]				& 25				& 15				& 10\\
$Y_{He}$  			& 0.10			& 0.10			& 0.11 \\
$\epsilon_{\rm C}$		& 7.00$\pm0.24$	& 7.00$\pm0.30$	& 7.00$\pm0.30$ \\
$\epsilon_{\rm N}$		&7.78$\pm0.23$	& 7.70$\pm0.20$	& 7.70$\pm0.20$ \\ 	
$\epsilon_{\rm O}$		&7.90$\pm0.20$	& 7.96$\pm0.30$	& 7.96$\pm0.30$\\ 
Fe/Fe$_{\odot}$		& 0.2				& 0.2				& 0.2 \\
$[\rm C/\rm H]$ (dex)	$^{\mathrm{a}}$ 		&-1.39			&-1.39			& -1.39 \\
$[\rm N/\rm H]$ (dex)$^{\mathrm{a}}$		& 0.00			&-0.08			& -0.08  \\
$[\rm O/\rm H]$ (dex)$^{\mathrm{a}}$  		&-0.76			&-0.70			& -0.70 \\	
\finf\					& 0.03			& 0.10			& 0.10 \\
\vinf\ 					& 2180.			&1300.			& 1400. \\
$\beta$				& 0.8				& 1.0				& 1.0  \\
log(\mdot$_{FUV}$) 		& -6.60			& -7.82{\bf $^{\mathrm{b}}$} 			& -7.95{\bf $^{\mathrm{b}}$}  \\
log(\mdot$_{Tramper}$) 	&-5.85			& -6.25			&-5.50 \\
log(\mdot$_{Vink}$) 		&-6.07			& -6.51			& -6.41 \\
log(\mdot$_{Lucy}$)		&-6.82			& -7.65			& -7.54\\
	   \hline
\end{tabular}
 \label{tab2}
\begin{list}{}{}
\item[] Uncertainties on \teff\ are $\pm$ 1000~K,  $\pm $ 0.2 dex for \logg, $\pm $ 0.1 dex on \logL. As for the wind quantities, 
an uncertainty of $\pm$ 0.2 dex was estimated for the mass-loss rates (given in \msunyr), while \vinf\ was measured within $\pm$ 100 \kms. 
For \mbox{CNO} abundances, $\epsilon_{\rm X}  =$ 12 +  log (X /H ).
\item $^{\mathrm{a}}$ Defined as  [X/H] =  log (X/H) -- log (X/H)$_{\odot}$. Solar values adopted from \cite{asplund05}. 
\item $^{\mathrm{b}}$ See Sect. \ref{sect_hydro} for a discussion about the true mass-loss rates from FUV
\end{list}
 \end{table}

The major modification with respect to \cite{tramper14} is about the wind parameters, that is the mass-loss rates, wind terminal velocities and clumping filling-factor, which we derive from the 
modeling of FUV P~Cygni profiles. 
Fitting the \cosp\ spectrum of IC1613-A13 required a volume filling factor of \finf\ = 0.03. 
Aside from the well-known clumping diagnostics such as the \ov\ \lb1376 and \niv\ \lb1718 lines, clumping turned out to be an important ingredient to fit the \civ\ \lb\lb1548, 1550 P~Cygni profile.
This is because the ionization balance of carbon is shifted toward C$^{4+}$ throughout the wind, at the effective temperature of IC1613-A13 and clumping helps C$^{4+}$ recombine 
to C$^{3+}$. More generally, this is true for  \teff\ in the range usually covered by O4 dwarfs and earlier (typically \teff\ $\ge$ 42,000 K). 
For \teff\ as high as quoted in \cite{tramper14}, the ionization fraction of C$^{4+}$ is actually close to unity and the resonance doublet is weak for an homogeneous wind. 
Significant level of clumping is required to produce enough C$^{3+}$ such that  a strong \civ\ \lb\lb1548, 1550 P~Cygni profile forms again. This dependance of the \civ\ profile on the clumping 
volume filling factor might explain why its more difficult to see the trend of \mdot\ with metallicity in A13 than in the late-type supergiants (cf. Sect \ref{sect_morpho}).

The clumped mass-loss rate is 2.5 $\times 10^{-7}$ \msunyr, while the wind acceleration is $\beta$ = 0.8 and the wind terminal velocity is 2180 \kms. The later are in excellent agreement with the values 
derived by \cite{garcia14} from the modeling of low resolution \cosp\ spectrum of this star, with the SEI method \citep[Sobolev with Exact Integration,][]{hamann81, lamers87}.


\subsection{IC 1613-B11 and WLM-A11}
\label{sect_res_a11b11}

The photospheric parameters derived for IC 1613-B11and WLM-A11 from the analysis of their FUV + optical spectra are consistent with their spectral classification, and
in relatively good agreement with those of \cite{tramper14}. 
The surface gravity we derive for IC 1613-B11 is actually more consistent with \logg\ derived by \cite{bresolin07}. 
We note that the \siiii\ lines (\lb\lb1295 -- 1299) get weaker as \logg\ decreases. They argue against \logg\ $\leq$ 3.2, which we adopted as a lower limit as it provides the best fit to the FUV spectrum while still maintaining a very good agreement with the optical spectrum. 
The uncertainty on \logg\ from the optical analysis alone is quite large because of the modest signal-to-noise ratio of the \xshooter\ spectra. Using the FUV spectra
in this case does help better constrain the value of the surface gravity.

We also note that, including error bars, WLM-A11 is over-luminous (by $\approx$ 0.2 -- 0.3dex), compared to luminosity class I stars with similar spectral type in the SMC or LMC 
\citep[see e.g.][]{massey09}. 
This trend is also present, although even stronger, in \cite{tramper11, tramper14}, and is consistent with the ``a'' subscript on its classification.

\begin{figure}
\centering
\includegraphics[scale=0.5, angle=0]{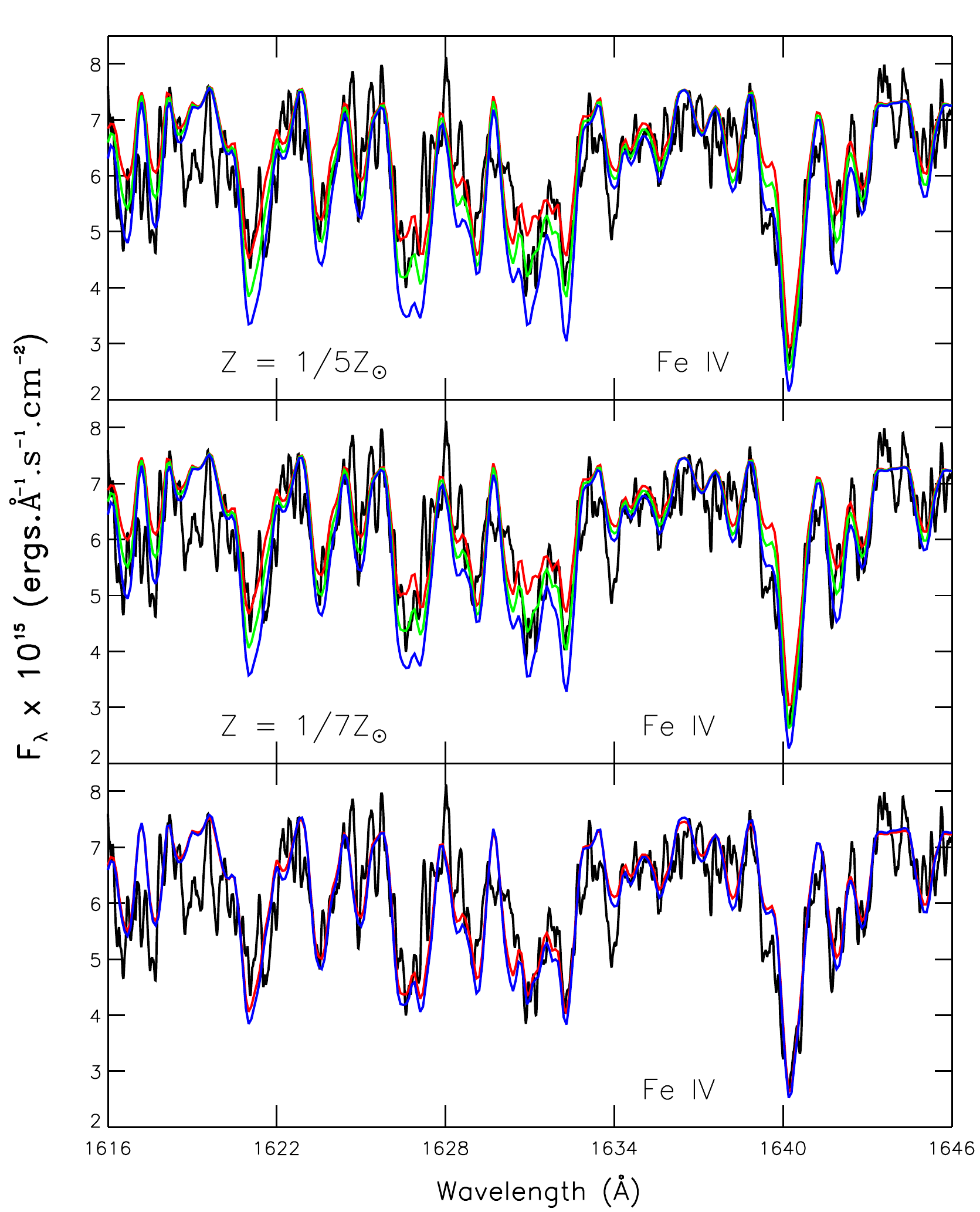}
\caption[]{In black, a portion of  observed spectrum of WLM-A11 showing several lines of \feiv\ around \heii\ \lb1640. 
The sensitivity of the \feiv\ lines tp the micro-turbulence is shown, for models having \xit\ = 7 \kms\ (red), 10 \kms\ (green) and 12 \kms (blue).  The top panel presents
models computed with $\fe/\fe_{\odot}=0.2$ (SMC-like), the models in the center panel have  $\fe/\fe_{\odot}=0.14$. The bottom panel shows a
models with $\fe/\fe_{\odot}=0.2$ and \xit\ = 10 \kms\ (red), and a model with $\fe/\fe_{\odot}=0.14$ and \xit\ = 13 \kms\ (blue).}
\label{fig4}
\end{figure}

For the photospheric parameters listed in Table \ref{tab2}, the iron lines in the FUV spectra of IC 1613-B11 and WLM-A11 are well reproduced with models having 
$\fe/\fe_{\odot}=0.14$ and \xit\ = 17 \kms\ and 13 \kms, respectively. 
However, minor changes to the micro-turbulence in the models (a decrease by 3 \kms) to increase Fe/H to SMC-like value provides as
good a fit as well. Figure \ref{fig4} illustrates the impact of such a minor variation of \xit\ on the profiles of \feiv\ and \fev\ photospheric lines.
Although the SNR of the \cosp\ spectra do not allow to rule out the lower metallicity ($\fe/\fe_{\odot}=0.14$) option,  we found that a SMC-like silicon abundance is more supported by the lines 
of \siiii\ (see Sect. \ref{sect_model}) present in these late-type O supergiants. 
The synthetic spectra of IC 1613-B11 and WLM-A11 presented in the Appendix  were obtained 
with models having an iron content $\fe/\fe_{\odot}=0.2$, which we adopted as the metallicity of IC1613-B11 and WLM-A11.

As expected from their evolved status, the surface abundances of IC 1613-B11 and WLM-A11 present signs of \mbox{CN} processing. Carbon is depleted
while we measured nitrogen enhancement with respect to the adopted baseline values for the metallicities of IC 1613 and WLM. 
On the other hand, the scarcity of sensitive diagnostics for oxygen abundance in the FUV makes its measure questionable for these two stars. In the optical, several lines of \oii\ are
present, in principle, but are very weak and are not useful given the moderate signal-to-noise ratios of the \xshooter\ spectra.  
We checked that a simple scaling to 0.2 the solar oxygen content gives the best agreement with the observed spectra however. With this scaling, the
sum of \mbox{CNO} scaled to 0.2 (as for the SMC) its solar value is also roughly conserved.
Finally, we find that helium is slightly enriched in WLM-A11, while no change to He/H is needed to reproduce \he\ line strengths in IC1613-B11. 
  
\begin{figure*}
\centering
\includegraphics[scale=0.5, angle=0]{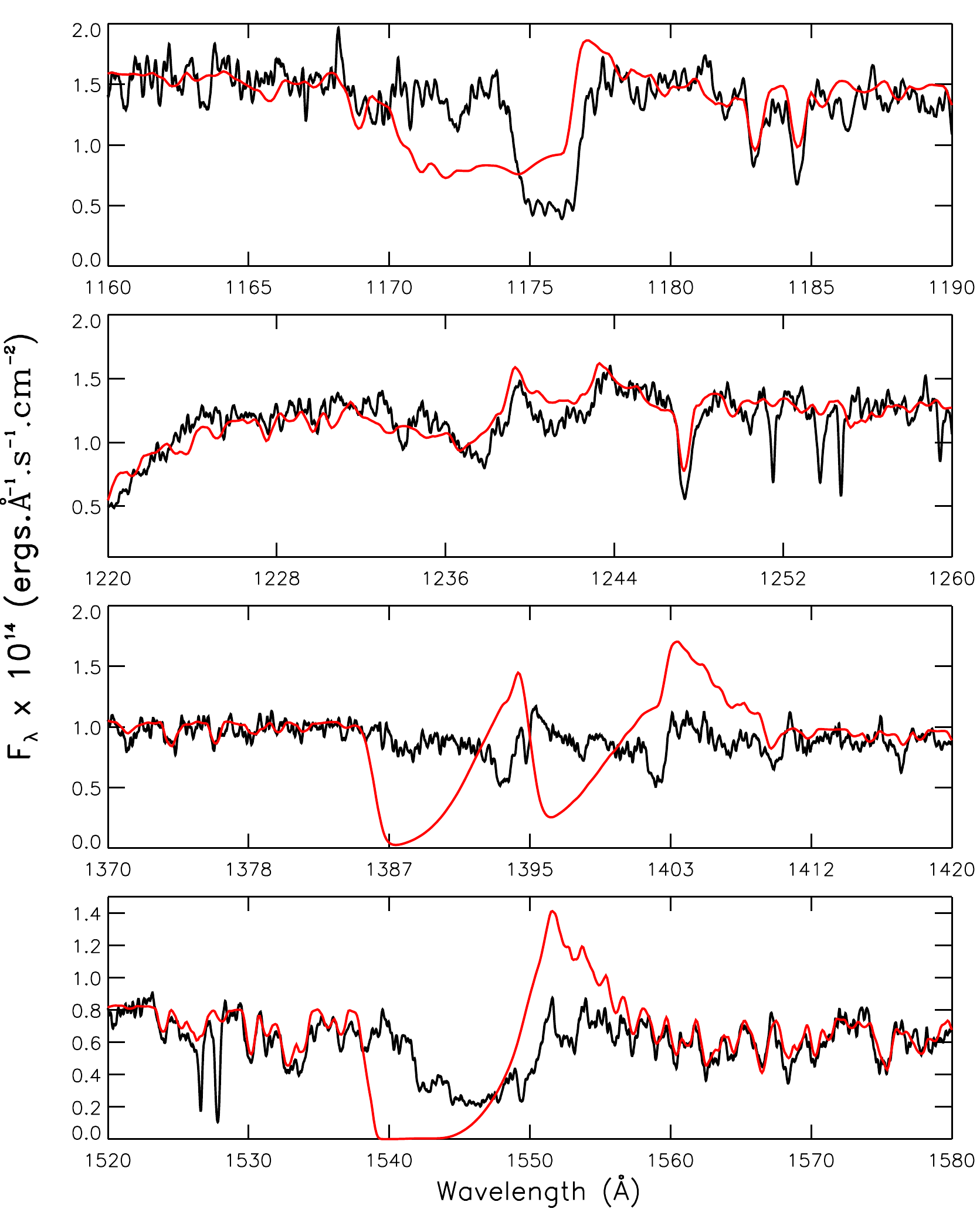}
\includegraphics[scale=0.5, angle=0]{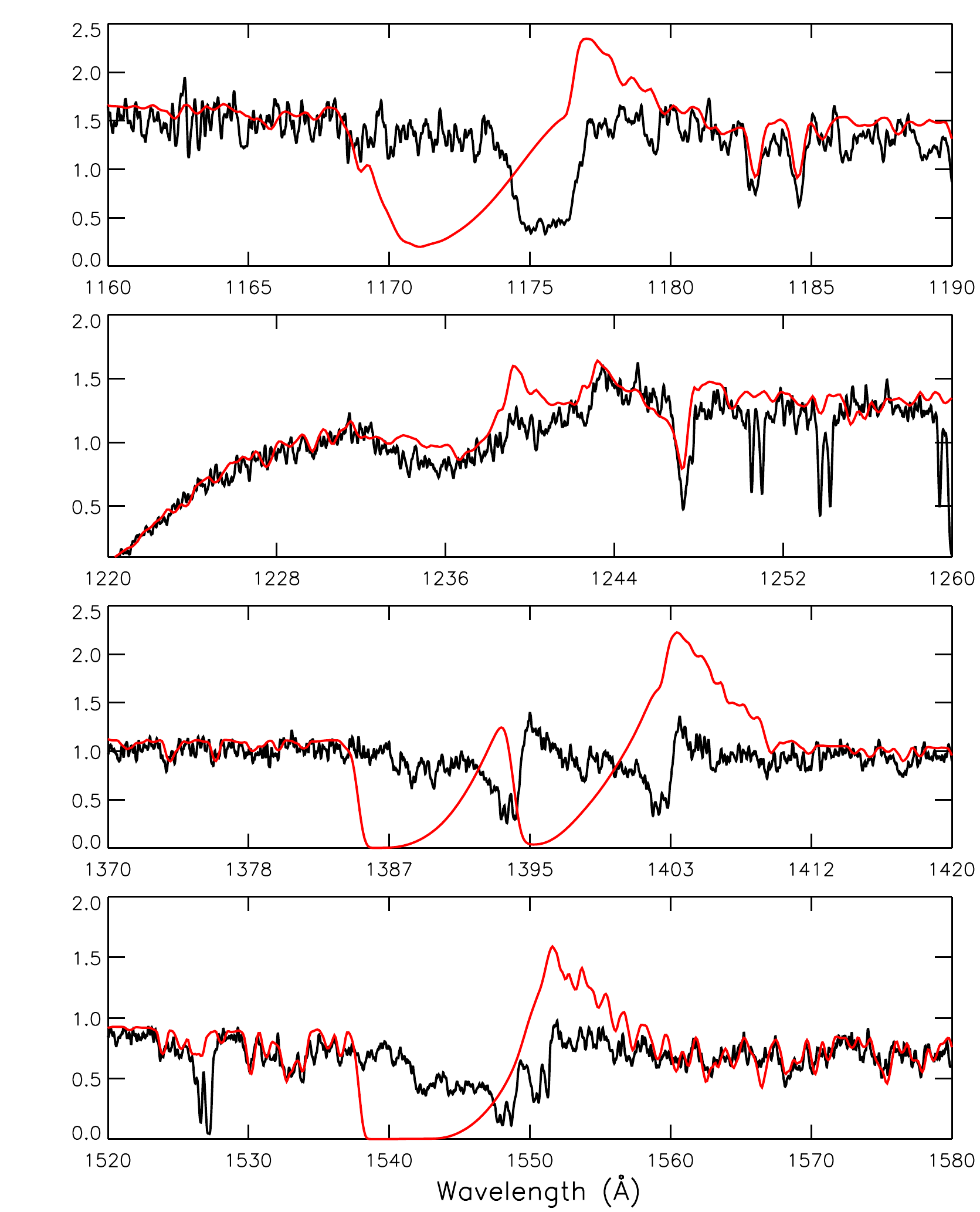}
\caption[]{In black, line profiles for different transitions in the FUV range, observed for IC1613-B11 (left) and WLM-A11(right). Overplotted in red, 
P~Cygni profiles produced by models using the parameters from \cite{tramper14}.
}
\label{fig5}
\end{figure*}

Here again, the most salient discrepancies with the study by \cite{tramper14} are the wind parameters. The high mass-loss rates they derived are firmly ruled out, producing much stronger P~Cygni profiles than observed (cf. Fig. \ref{fig5}). In our modeling, the (clumped) mass-loss rates are 1.5$\times 10^{-8}$ \msunyr and 1.1$\times 10^{-8}$ \msunyr\ for IC 1613-B11 and WLM-A11, respectively. 

With these models, we could obtain a reasonable agreement with the observed P~Cygni profile of \civ\ resonance doublet, the \siiv\ resonance doublet as well as \nv\ \lb\lb1238, 1242, although 
none of our models could fit these three profiles with the same level of quality simultaneously. The case of IC1613-B11 is the most problematic as we could never reach an excellent fit
of the \civ\ profile for this star with our models. At low velocities, the absorption component of the observed P~Cygni profile is stronger than predicted by the models, unless we increase \mdot\ to such extent that in the synthetic profile, the absorption is too strong near \vinf\ and we also produce a strong emission component that is not observed in \civ. Note that this behavior is 
typical of late-type O dwarfs exhibiting the "\it weak-wind\rm" problem, although IC 1613-B11 is above the luminosity threshold usually found for these stars 
\citep[\logL\ $\approx$ 5.2, e.g.][]{martins05b,marcolino09}.
 Furthermore, a well-defined P~Cygni profile is  produced for the \siiv\ doublet, also not observed. 
 
The P~Cygni profiles, although relatively weak, presented good sensitivity to variations of the clumping filling-factor. 
The filling factor is moderate in both cases, that is \finf\ = 0.1, implying a reduction of a factor of three for \mdot\ compared to an homogeneous case.
The \mdot\ we derive correspond to a regime where H$\alpha$ is mostly insensitive to constrain mass-loss rates \citep{marcolino09}. 
The modeling of the FUV and optical spectrum of supergiant stars in the SMC (Bouret et al., in preparation) yields mass-loss rates in good qualitative and quantitative consistency with those derived 
here for IC 1613-B11 and WLM-A11. The mass-loss rate derived for AzV 327, for instance, is roughly three times as high as those of IC 1613-B11 and WLM-A11, a trend that is anticipated from the 
relative strength of their P~Cygni profiles (cf. Fig. \ref{fig1}). 

The exponent of the velocity law was constrained from the weak emission component of the P-Cygni profile. Its value is consistent 
with theoretical predictions \citep{muijres12}, but is lower than found by \cite{mokiem06} or \cite{massey09} for late-type supergiants. 
Furthermore, the wind terminal velocities are consistent with \vinf\ measured by \cite{massey09} from UV data for late-type supergiants in the Magellanic Clouds.
However, we stress that the P~Cygni profiles in B11 and A11 are weak enough that the wind terminal velocities are rather difficult to measure. 
Most particularly, several lines of \feiv\ overlap the blue edge of the \civ\ resonance doublet P~Cygni profiles. These lines were not an issue for IC 1613-A13 but they are stronger at the 
lower \teff\ of B11 and A11. With $\Delta$V $\approx$ 1200 \kms, they fall
right in the typical range for \vinf\ indicated for B11 and A11 by the discrete absorption components, following \cite{prinja98}.
Although the \cmfgen\ models do account for these \feiv\ lines, their sensitivity to \teff, the adopted Fe/H, the microturbulence and even \logg\ could introduce an additional uncertainty in the determination of \vinf, with expected repercussions on the predicted mass-loss rates (see Sect. \ref{sect_pred_wind_prop}). 
%
%

\section{Metallicity of the massive stars of IC 1613 and WLM} 
\label{sect_metal}
The striking conclusion by \cite{tramper11} of a discrepancy between observed and predicted mass-loss rates 
relied for a large part on the 
adopted metallicities for IC 1613,  WLM and NGC 3109. The (low) metal content for these galaxies has been measured by various techniques, which all point towards a SMC-type metallicity or even below. 
In their more recent work, \cite{tramper14} acknowledged that the metallicities they considered in their initial study
were likely underestimated. In the present study, we have a direct access to the global metallicity of the stars through thousands of photospheric lines of iron ions 
(\feiii\ to \fevi) present in the FUV spectral range. Oxygen abundance diagnostic is also available (cf. Sect. \ref{param_sect}).
Before comparing empirical mass-loss rates as derived from the modeling of FUV spectra to
theoretical ones, we first need to discuss how the metal content we derived compares to previous determinations.

\subsection{Oxygen}
\label{sect_oxy}
As discussed in Sect. \ref{sect_res_a13} and \ref{sect_res_a11b11}, the oxygen abundance is difficult to constrain accurately in the three stars. The
best we can say at this point is that oxygen abundances such that the total \mbox{CNO} is conserved do provide good fits to the scarce diagnostics available. In principle,  the 
evolved status of IC1613-B11 and WLM-A11 should translate into oxygen depletion, compared to the initial \mbox{CNO} mixture. Although we do not need such a depletion to reproduce the 
\oiv\ lines present in the FUV spectra of both stars, we cannot rule out that oxygen abundances are actually lower. Only for IC 1613-A13 does such a depleted oxygen content manifest itself (maybe a consequence of the earlier spectral type, hence higher mass, implying faster evolution), but is however moderate. 
In summary, O/H in the three stars ranges between  1/7 -- 1/5 the solar oxygen content, but we emphasize again that these estimates are poorly constrained and may be revised provided stronger
diagnotiscs become available. How relevant these abundances are when dealing with comparisons between observed and predicted mass-loss rates of massive stars 
is however questionable. 

\subsection{Stellar Iron Content}
\label{sect_iron}
Although the low oxygen (and more generally $\alpha$-elements) content of IC 1613 and WLM is confirmed from FUV spectroscopy, direct measures of the iron abundance are more relevant to study the metallicity dependance of mass-loss. It is expected that the mass-loss rates of massive stars are determined by the the radiative acceleration below the sonic/critical point in the wind, 
which is predicted to be driven by the iron group elements for O-type stars in the metallicity regime of our sample \citep{vink01}. Therefore, such measures of the iron abundance as those we achieve in this work are mandatory to compare empirical mass-loss rates to theoretical predictions, and in any case are more relevant than metallicities based on $\alpha$-elements abundances. 
Iron abundance determination of course relies on the strength of iron lines which depends on the actual iron ionization. The later depends on the effective temperature and gravity \citep[e.g.][]{heap06}. For a given temperature, the ionization is higher at lower gravities as expected from the Saha equation and the numerous iron lines grow stronger. Iron lines may also look stronger because of higher microturbulence \xit, as shown in \cite{hillier03} and \cite{ bouret03}. This degeneracy of the ionization balance of iron with the photospheric parameters is such that deriving the iron abundance from fits
to the iron forest requires \teff, \logg\ and \xit\ to be determined from other diagnostics, which we have at our disposal through the analysis of the optical spectra for instance.
From the joint analysis of the FUV and optical data, we found that for the three stars presented here, [Fe/H] = -0.66, i.e. their iron content is SMC-like. 

In IC 1613, stellar iron abundances of [Fe/H] = -0.67 have been inferred from spectroscopy but for late-type, evolved stars \citep[e.g. M type supergiants][]{tautvaisienne07}.
Alternatively, photometric studies were performed to measure the iron to hydrogen ratio of stars on the red giant branch \citep{cole99, Skillman03, minniti97, McConnachie05}. 
These measures, however, determine the iron content of an older stellar population, while the objects we are dealing with here belong to a young stellar population. 
To our knowledge, no other measures exist of the present-day iron abundance as probed here through high spectral-resolution FUV spectroscopy and quantitative analysis of the forest of iron lines. 
The iron content we derive for the two stars in IC 1613 is also compatible with the study of the star formation history of IC 1613 by \cite{Skillman03}; 
these authors indeed found that the metallicity increased from [Fe/H] = -1.3 to -0.7 over the age of the galaxy.
It is also noteworthy that using oxygen $+$ silicon abundances as a proxy for the global abundance of $\alpha$-elements, [$\alpha$/\fe] is solar
in our case (or slightly lower for IC 1613-A13). Given the uncertainties in both the abundances for $\alpha$-elements as well as iron,  this is however in qualitative agreement with the 
conclusions of \cite{tautvaisienne07} for IC 1613. Finally, we note that from a qualitative comparison of FUV spectra of OB stars in IC 1613 with OB stars in the SMC,
\cite{garcia14} also argued that the iron contents are similar in both galaxies. 

On the other hand, \cite{Urbaneja08} quote a (present-day) metallicity of log\,(Z/Z$_{\odot}$)=-0.8 ($\pm0.2$) for WLM-A11, although it is not derived from the direct analysis of iron-group lines. 
Although this value for Z/Z$_{\odot}$ agrees with our estimate of the metallicity within the error bars, the average metallicity of the A and B supergiants analyzed by \cite{Urbaneja08} is  log\,(Z/Z$_{\odot}$)=-0.87 ($\pm0.06$). This lower value is not favored by the spectroscopic analysis in our case, although better signal-to-noise ratio of the spectrum of WLM-A11 would be required to firmly discard this solution. We note that the metallicity of A-supergiants analyzed in WLM  by \cite{venn03} is however even higher than ours (with [Fe/H] = -0.38 $\pm 0.20$). The ranges in [Fe/H] 
are large enough that despite measure uncertainties, they hint to a real scatter in the star-to-star metal content in WLM. This comes as a surprise as WLM-A11 and the stars analyzed in \cite{venn03}
are located within the same region of the WLM galaxy.

Overall, our joint analysis of the FUV and optical spectra reveals that the iron content of the three stars is similar to the SMC. In this sense, the theoretical mass-loss rates are expected to be higher than previously considered, 
although by a fairly limited factor of $\approx$ 1.3. This is far from being enough to account for the strong discrepancy discovered by \cite{tramper11} and
confirmed in \cite{tramper14}, 
especially for the late supergiants. The wind terminal velocities are also impacted by the metallicity, although to lesser extent, which does not change this conclusion. 


\section{Binary status}
\label{sect_binary}
Multiplicity can impact the mass-loss determination in several way \citep[see the discussion in][]{tramper11}. The issue of multiplicity among massive stars has become very important 
in the past few years, with results for galactic massive stars pointing to fractions of stars with companions ranging from 75\% \citep[][for clusters and associations in this specific case]{mason09} 
and up to a striking 91\% \citep{sana14}. At lower metallicity, results obtained within the VLT-FLAMES Tarantula Survey \citep[][and refs therein]{sana12a}, point to a fraction of 
more than 50\% binaries within the sample, once corrected for observational biases. 

Concerning IC 1613-A13, the SED fitting is compatible with the star being single; the fluxes from the \cosp\ spectrum and the photometry
(cf. Table \ref{tab1}) are well accounted for by the SED produced by a single-star model as computed with \cmfgen. 
The luminosity we derive is slightly higher (within 0.2dex) than those we obtained for (three) stars with close spectral type and effective temperature in the SMC \citep{bouret13}. 
However, one of these stars, namely MPG 324 that we used for morphological comparison in Sect. \ref{sect_morpho} (see also Table \ref{tab2} and Fig. \ref{fig2}) is probably a single-lined binary as
indicated from small radial-velocity variations \citep[]{evans06}, even though its theoretical SED from \cmfgen\ is compatible with the observed one.
The case of MPG 324 suggests that we cannot firmly rule out the presence of a companion for IC 1613-A13 from the SED fitting approach only.  Nevertheless, we can safely 
conclude that since the putative companion is of low mass enough that it does not show up in the observational SED, it is unlikely to impact the wind signatures we observe in the
FUV spectrum.   

The effective temperature, surface gravity and luminosity of IC 1613-B11 are compatible with those derived for stars with same spectral type/luminosity class in the MCs \citep[e.g.][]{massey09}, for
which no signs of multiplicity are known. Much like for IC 1613-A13, we do not see an influence from a potential companion on the SED of IC 1613-B11;  here again, if the star is in a binary system, 
it must be with a companions of late type, whose luminosities must be low, hence, expectedly, the mass-loss rate.
More significantly, the photospheric parameters of IC 1613-B11 are consistent within the error bars, with the calibration in \cite{martins05} for an O9.5 supergiant. 
Although low metallicity stars are known to be on average hotter and more luminous than their galactic counterparts for a given spectral,  it is well-known that theses differences 
decrease at later spectral type and practically vanishes by B0 \citep[see e.g. Table 9 in][]{massey05}. The spectroscopic mass is also consistent with this calibration, 
which further supports the single star interpretation.    

The case of WLM-A11 is somewhat different. The SED fitting indicates that this star is  more luminous than a normal O9.7 supergiant (also denoted by the "a"
subscript of its spectral classification). The photospheric radius we derive and the surface gravity we measure are such that the star is then roughly twice as massive as
other late-type supergiants of similar spectral properties, including IC 1613-B11 or stars in \cite{massey09}.
Such a  high spectroscopic mass ($\geq 53$ \msun) is actually more characteristic of O-supergiants with early spectral types \citep[e.g.][]{massey09, bouret12}.
This might indicate that WLM-A11 is actually in a binary system or that there is a unresolved star in the fore/back-ground, contributing to the observed fluxes.  
We will come back on the influence of changing the stellar luminosity on mass-loss prediction, in a following section. 

\section{FUV mass-loss rates and uncertainties}
\label{sect_error}
\cite{tramper11, tramper14} relied on the fitting of the H$\alpha$ (and \heii\ \lb4686 line for the hotter objects) to measure the mass-loss of their targets.
Known advantages are an easy and robust abundance determination and ionisation balance calculation. The sensitivity of
H$\alpha$ is limited to stronger winds however, showing only moderate line core filling for mass loss rates between $10^{-6}$ and $10^{-7}$ \msunyr\
\citep[see Fig. 5. in][]{marcolino09}. Because of their low metallicity, our targets should have mass loss rates in this parameter range \citep{vink01}. 
Detailed investigations should therefore rely on diagnostics showing sensitivity in this regime, namely FUV. Obviously, mass loss determinations in this
spectral range also suffer from specific uncertainties related to the assumptions adopted in the modeling. 

\subsection{Abundances, Ionization and Wind Acceleration}
\label{sect_ab_ion}
The observed \cosp\ spectra present three important lines of carbon ions, namely \civ\ \lb1169, \ciii\ \lb1176 and \ciii\ \lb1247, that we used to measure
the photospheric  abundance of carbon. Within the uncertainties on C/H, the mass-loss rates of the targets could vary by at most a factor of two (but see below
for a comment on the influence of \mdot\ on the  \ciii\ \lb1176 line in the late-type supergiants). 
We emphasize that the predicted (wind) profiles of the \siiv\ resonance doublet are reasonably reproduced in our models of the late supergiants (cf. Fig. \ref{fig_a3} and Fig \ref{fig_a5}). 
The \mbox{Si} element does not
suffer, a priori, from the uncertainties related to carbon, as it is not involved in nuclear burning until very much later phases of the evolution of massive stars. Changing \mdot\ by a factor of two
(i.e. within the boundaries of the relative uncertainty due to the carbon abundance) would change the profiles of \siiv\ resonance doublet enough that we can rule out these solutions. 

The other obvious uncertainty is related to the ionization fraction of \mbox{CNO} ions. This might be a potential flaw of the FUV-based mass-loss determination,
compared to hydrogen-helium based methods for photosphere $+$ wind properties determination.
This is because of uncertainties in the predicted flux levels at $200 \leq\ \lambda \leq\ 400$ \AA, or more exactly differences in the flux levels predicted
by state-of-the-art NLTE, line-blanketed models for massive stars \citep{puls05}. These differences  translate in differences in the predicted populations of important ions such 
as \civ\ or \nv\ for instance, hence on the estimate of the mass-loss rate. 
\cite{puls05} emphasized that the uncertainties on the flux levels mostly impact the ionization fractions in the outer part of the winds, 
where they translates into differences in population of typically $\pm$ 0.15 dex, while they agree almost perfectly everywhere else.
In our case, we are able to reproduce line ratios of different ionization stages of carbon, nitrogen, and oxygen which should guarantee that the flux slope in the critical
range $200 \leq\ \lambda \leq\ 400$ \AA\ is correct. In any case, the ionization fraction of carbon, nitrogen and silicium ions should be accurate within a factor of
a two, which does not change our conclusion on the actual mass-loss rates of our targets.  

We can obtain a good grip on the wind velocity law, acceleration parameter and terminal velocity, from fitting of the slope of the emission component of the P~Cygni profiles. 
It is in any case more robust than what can be done from optical lines alone when the line profiles are in absorption, in which case these parameters
are not accessible thus increasing the uncertainty in the derived mass-loss rate, by a factor from 2 to 3 \citep{puls96}. 
In our case, the uncertainty in $\beta$ and \vinf\ reflects in a moderate correction on \mdot\ only, which again does not change the 
conclusion of the present paper. 
Note that  uncertainties on \vinf\ have an influence when dealing with the modified wind-momentum-luminosity relation \citep[see e.g.][]{kudritzki00}. 
The values of \vinf\ we measure are for instance $\approx$ 25\%\ different from those 
listed in \cite{tramper14}, higher for IC1613-A13, but smaller for IC1613-B11 and WLM-A11. 

\subsection{Clumping}
\label{sect_clumping}
Wind clumping and the formalism used to describe it, is another important issue for the determination of mass-loss rates. It modifies the predicted line profiles for a given \mdot\ 
and this parameter cannot be considered separately of clumping parameters anymore. 

Clumping as implemented in \cmfgen\ is usually referred to as micro-clumping \citep[][and refs therein]{hillier99}. Its influence on \mdot\ determinations in O stars, both 
from FUV and optical diagnostics has been tested and discussed extensively for stars in the Galaxy and the SMC \cite[e.g.][]{hillier03, bouret05, bouret12, bouret13}. 
The volume filling factors (\finf) associated with the clumped models for our three targets (\finf\ = 0.03 for IC16-A13, and \finf\ = 0.1 for IC1613-B11 and WLM-A11) 
are in line with those found in these studies. They imply that the mass-loss rates are scaled down by factors 6 and 3 (respectively), compared to the homogeneous values that would be derived from fitting H$\alpha$. 
Note that with the adopted parameterization of clumping in\cmfgen\ (see Sect. \ref{sect_model}), the volume filling factor in the formation region for the H$\alpha$ line is much higher than the value \finf. 

However, the filling factor approach is admittedly too simple. It considers optically thin clumps, an assumption valid for the continuum but not so much for the lines. Furthermore, it does not account for the interclump medium which is important for producing the P~Cygni profiles of super ionized species \citep{zsargo08} or, even more fundamentally, because void interclump medium is physically unlikely \citep[e.g.][]{owocki88, surlan13, sundqvist14}. 
It also ignores porosity, both in spatial and in velocity space \citep[e.g.][]{oskinova07, sundqvist11}. 
A detailed study by \cite{sundqvist14} concluded that velocity-porosity, \emph{a.k.a vorosity}, impacts both the line profiles (diagnostics) and mass-loss properties (dynamics),
in contradiction with an earlier result by \cite{surlan13} and based on a different approach, showing that spatial porosity was most important. 

Notwithstanding the very different levels of physical sophistication to describe the nature of wind clumping and its effets (e.g. the role of porosity/vorosity), studies have found that the 
corrections to mass-loss with respect to theoretical predictions \citep[the widely used recipe by][]{vink01} should be less than a factor of 1.5 -- 2.5 typically 
\citep{bouret12, surlan13, sundqvist14}. This difference is actually very reasonable given the uncertainties related to both approaches, namely NLTE radiative transfer codes 
used to describe clumped stellar winds on one hand, and Monte-Carlo calculations of the radiative force used to predict mass-loss rates on the other hand.

Although these results were obtained for the same early-type Galactic supergiants with fairly dense winds, it is tempting to argue that the mass-loss rate we obtained, 
at least for IC1613-A13 which has a relatively dense wind, is realistic within the same margin, enough to provide a 
robust comparison basis against theoretical predictions.

\subsection{Line force, momentum equation and implications}
\label{sect_hydro}
Whatever the physical nature of clumping, the presence of inhomogeneities in the wind makes it very difficulty to determine the systematic errors associated with our mass-loss rates estimates.  
This is dictated by the lack of diagnostics, a problem exacerbated in stars that have weaker winds. Typically in the FUV we only have a few lines suitable for mass-loss determinations, and these line often have other sensitivities.  For example in A13, the only wind diagnostics are the \civ\  and \nv\ doublets, \niv\ \lb1718, and the absence of \ov\ \lb1371, while in A11 we loose \ov\ as a 
diagnostic but gain the \siiv\ doublet. For an inhomogeneous wind we have multiple parameters that describe the the clumping, porosity, and porosity in velocity space that need to be determined.

In our modeling we have adopted the volume filling factor approach which has only two free-parameters, and one of which (the location at which clumping starts) is not well constrained. However, we have one additional consistency check that can be utilized. While \cmfgen\ does not compute the mass-loss rate self consistently, it does compute the radiative force, hence we can check how well the momentum equation is satisfied.

For A13 we find that the momentum equation is not satisfied -- we deposit too much momentum in most of the wind, and this inconsistency cannot be changed by altering the velocity law while remaining constant with the observations\footnote{We ignore issues in accelerating the wind around the sonic point. In general we have too little line force at, and just above, the sonic point to accelerate the flow. Much of this discrepancy arises from issues with the choice of the microturbulent velocity: driving a stationary wind through the sonic point requires the microturbulent velocity to be much less than the sound speed. Although this is typical for O supergiants, it is not the case here --  the microturbulent velocity is of order 0.5 times the sound speed.}.
Three possible solutions to the inconsistency are:
\begin{enumerate}
\item
An incorrect ionization structure.
\item
Additional hot gas that we are not detecting. In the standard wind-driven wind model, it is assumed that hot shocked gas, arising from the radiation driven wind instability, rapidly cools. While this is likely to be valid for the wind of a star like $\zeta$ Pup \citep[see, e.g., discussion by][]{hillier93}, the assumption will eventually become invalid as the density declines
\citep{drew94, martins05b, lucy12}. Recently \cite{huenemoerder12} showed that the X-ray emission from $\mu$ Columbae originated in the wind, and required a hot gas component with a mass-loss rate in excess of that inferred from analysis of UV lines.
\item
We are underestimating the mass-loss rate due to the effects of porosity and/or vorosity.
\end{enumerate}
 
We dismiss the first one since in \cmfgen\ the predicted and observed ionization are consistent (see comments above).  For A13 we also argue that in its relatively dense wind, the hot gas component is probably not important. However, consistent with other findings, porosity and vorosity could be important. If we adopt the same \finf\ and \mdot = $5.5 \times 10^{-7}$ \msunyr\ we find much better agreement with the radiation force and that required to derive the wind. However, absorption features in the spectrum are now too strong.  For \finf\ = 0.1 and \mdot\ = $4.0 \times 10^{-7}$ \msunyr\ the situation is somewhat more intermediate -- we somewhat deposit too much momentum between 100 and 1600 \kms, and somewhat too little in the outermost regions ($V >$ 2000\,\kms). 

Thus for A13 we conclude that our derived mass-loss rate is underestimated, at most, by a factor of
two. This is consistent with our finding in $\zeta Pup$ where we found \mdot= $2.0 \times 10^{-6}$ \msunyr\  for
\finf\ = 0.05 using multiple diagnostics. However we could only fit the \pv\ resonance doublet with an abundance a factor of two less than solar. If we assume the solar abundance is applicable, this would imply that vorosity/porosity 
effects influence mass-loss rates derived from the \pv\ doublet by about a factor of two.\\

For A11 and B11, with much weaker winds, the situation is much more problematical.
Let us discuss the case of A11 as an example. The mass-loss rate of $1.1 \times 10^{-8}$ \msunyr\ is a factor of three less than the 
limit for a single optically thick line located at the emission peak of the SED $(L/c^2)$. 
This inconsistency is borne out when one examines the momentum equation -- the radiation force that we
derive is typically a factor of 5 to 10 times large than that required to drive the flow. This discrepancy indicates that some fundamental assumption in our modeling is incorrect.

For A11 we believe that the problem is twofold. In its weak wind it is possible that the gas does not
cool after it is shocked. Thus a significant fraction of the wind is in hot gas -- a result consistent with the
use of X-ray filling factors near unity in order to get a log$\rm L_{X}/L_{bol} \approx -7$. The second is that in these
weaker winds, the influence of porosity/vorosity is more important. An alternative is that wind dynamics is
influenced by the presence of a strong magnetic fields but since the issue of weak winds in persistent,
even in stars where a magnetic field is not detected \citep[e.g., 10 Lac,][]{david-uraz14}, this cannot be the answer. If
we adopt \finf\ = 0.1 and \mdot\ = $3 \times 10^{-7}$ \msunyr, there is reasonable consistency between
the radiative force and that required to drive the flow. Of course this mass-loss rate produces a spectrum
very inconsistent with that observed.

Independent of the cause of the discrepancy, we might expect our assumptions to be more
valid deeper in the wind. Thus  we decided to consider a model in which we varied the mass-loss
rate as a function of the radius $r$, thus mimicking a wind in which, for example, a lot of the gas becomes hot
and does not contribute to the observed UV spectrum. In the test case we assumed \mdot = $7.7 \times 10^{-8}$ \msunyr\ in the inner wind, and $1.1 \times 10^{-8}$ \msunyr\ in the outer wind, with a smooth transition between them. 
With this model two distinct features were revealed:
\begin{enumerate}
\item
The higher mass-loss-rate model desaturates the \ciii\ triplet line near 1176 \AA, and produces
a weak blue wing. The profile is in better agreement with observations, than our best fit model.
\item
The higher mass-loss-rate model produces a slight blueshift, and a slight broadening, of
the absorption associated with the \siiv\ resonance doublet. This blueshift and broadening
is also seen in the \ciii\ triplet.
\end{enumerate}

\begin{figure}
\centering
\includegraphics[scale=0.5, angle=0]{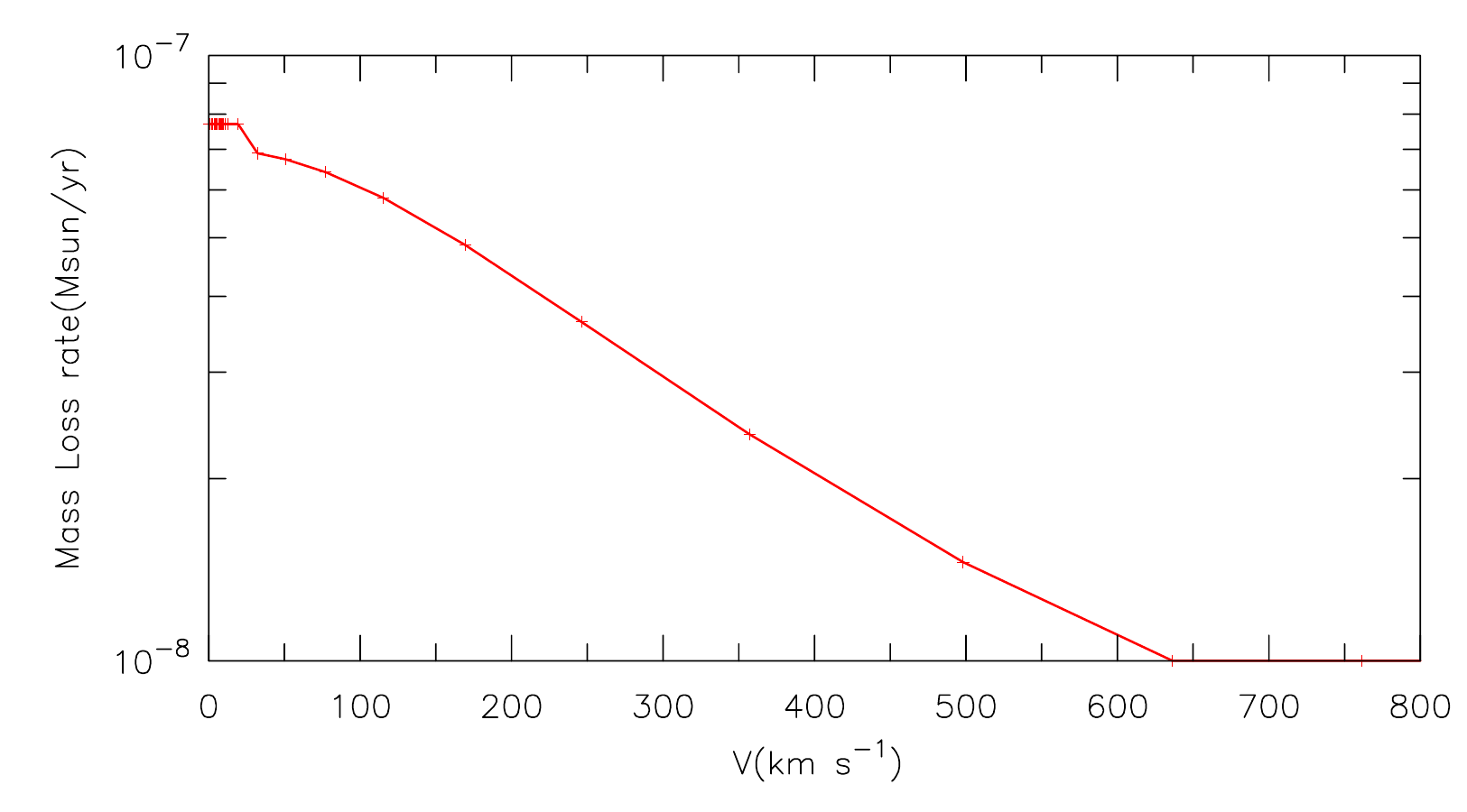}
\caption[]{Variation of the mass-loss rate, plotted here as a function of the velocity in the wind. The model was used to compute the spectrum presented in Fig. \ref{fig7} (blue line).}
\label{fig6}
\end{figure}

\begin{figure}
\centering
\includegraphics[scale=0.5, angle=0]{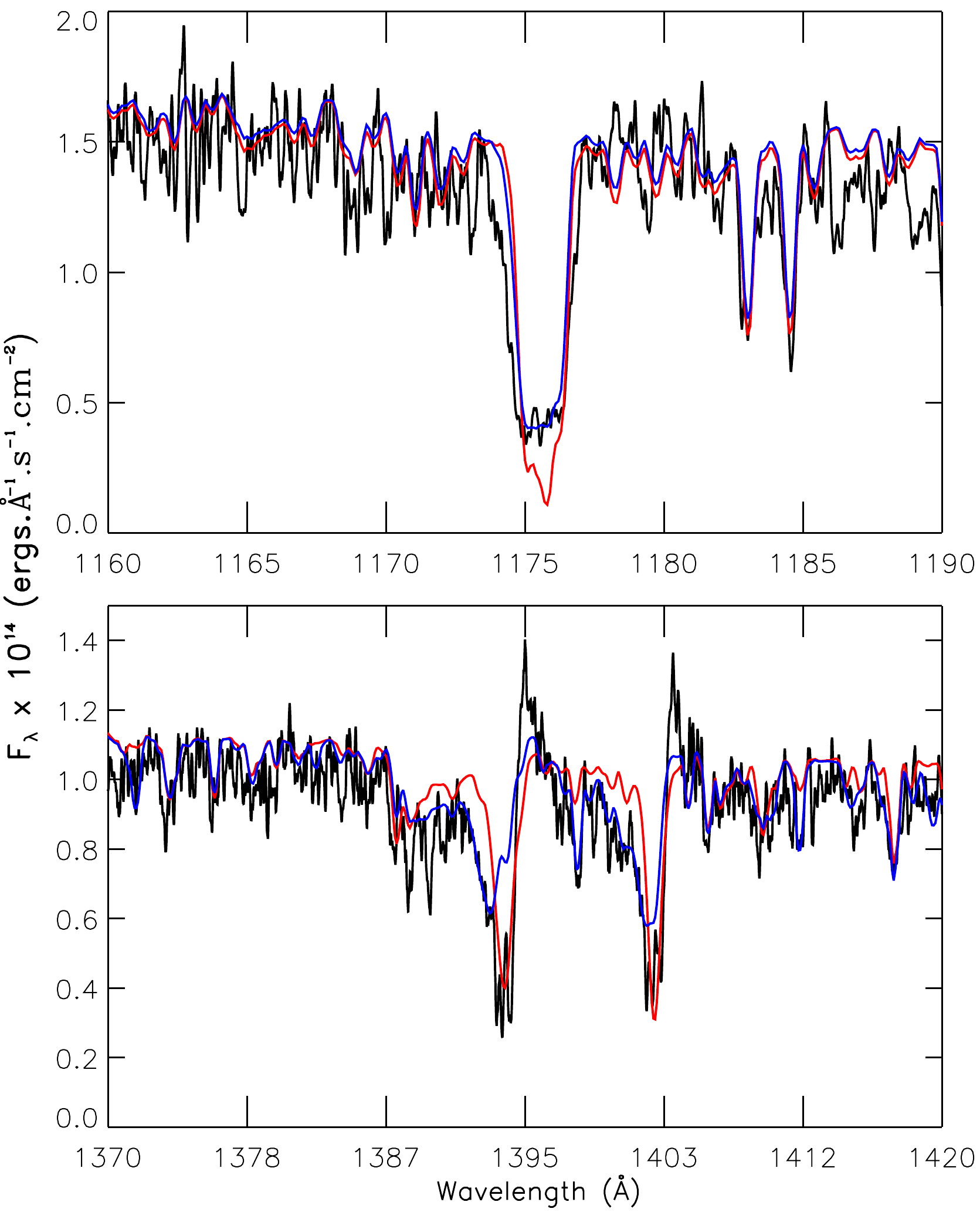}
\caption[]{Profiles of \ciii\ \lb1176 (top) and \siiv\ resonance doublet (bottom). The model with \mdot\ varying with radius (see text and Fig. \ref{fig6}) is plotted
in blue. Our best fit model is shown in red.}
\label{fig7}
\end{figure}

We thus argue the our mass-loss rate for A11 derived from the conventional lines is probably too low,
by some factor between 3 and 10. Again, similar conclusions hold for B11. In these weak wind stars, the influence of porosity/vorosity effects, 
and the possible presence of hot gas, can potentially severely bias the estimated mass-loss rate. 

\section{Wind properties at low metallicity}
\label{sect_obs_wind_prop}
From the modeling of the optical spectra of early-type stars, in three metal-deficient galaxies, more precisely galaxies with O/H less than SMC, \cite{tramper11, tramper14} 
strikingly concluded that the mass-loss rates are consistent with those that would be predicted by the radiatively-driven wind theory for a higher, LMC-type, metal abundance. 
This lead them to argue that  this theory would break down at sub-SMC metallicity. 
If confirmed, this would be a major problem for our understanding of the physics of stellar winds, and for stellar evolution in the upper HRD as well. 
However, several facts suggest that the conclusion by \cite{tramper11, tramper14} may be premature.

\begin{itemize}
\item 
We showed in Sect. \ref{sect_metal} that the iron abundance in IC 1613 and WLM is higher than the metallicity adopted from O/H \citep[][and refs therein]{tramper11}. In the
three stars, the metal content is not, if at all, significantly smaller than in the SMC, for which theoretical \mdot\ were found to match the empirical ones \citep{mokiem07} . 
The higher metal content shall impact the theoretical mass-loss rates as well as the wind terminal velocities, hence the modified wind-momentum-luminosity relation (WLR)  for the 
stars\footnote{This relation links the modified wind momentum, defined as \dmom\ = \mdot\,\vinf\,$\sqrt{R/R_{\odot}}$,  to the stellar luminosity. The WLR is often used to study mass-loss in various galaxies and environments \cite[e.g.][]{mokiem07} as \dmom\  is almost independent of mass.}
\item No important change in wind driving (e.g. going from a driving through lines of iron group elements to \mbox{CNO}) is expected until the metallicity is 1/50Z$_{\odot}$ and below \citep{vink01}. 
\item \cite{herrero12} found that the wind parameters of an O6.5 IIIf star in IC 1613 are such that the star
follows the WLR for its metallicity. In other words, if the metallicity would be sub-SMC \cite[but see the recent conclusions by][besides ours]{garcia14}, there is at least one case where the standard theory still applies. 
\item The sample analyzed by \cite{tramper14} contains 10 stars, a majority of them having wind strengths that are compatible with theory within the errors, assuming a SMC-like metal
content. Only three stars (all of them in NGC 3109) are real outliers in the WLR diagram. 
\item The intrinsic dispersion in the wind parameters within a given spectral/luminosity class typically ranges within a factor of two. This is of the same amplitude as the difference that is expected for two stars with the same parameters but a factor of 2.5 in the metal content (i.e. going from LMC to SMC-type metallicity). 
\end{itemize}

We showed in Sect. \ref{sect_res_a13} and \ref{sect_res_a11b11} that the FUV-spectra of IC 1613-A13, IC 1613-B11 and WLM-A11 indicate the presence of significant
wind clumping. All other things being equal, the mass-loss rates quoted in \cite{tramper14} should therefore be reduced accordingly.
The amplitude of the reduction depends on the nature of clumping (cf. Sect. \ref{sect_clumping}); the typical reduction of \mdot\ found for clumped versus homogeneous models 
is found to be a factor 3 to 6. 
Clumped mass-loss rates are reduced by less than a factor of three compared to theoretical predictions \citep{bouret12, surlan13, sundqvist14}. 
Note however that this last conclusion holds as long as the impact of clumping on the wind driving is not considered \citep[see e.g.][for comments]{muijres11}.
Recently, \cite{sundqvist14} found that when applying a vorosity correction to the line force \citep[based on the 
standard theory of line-driven winds by][]{castor75}, a reduction of the line force that drives the wind is obtained. Therefore, a reduction of theoretical mass-loss
rates is expected from these calculations if vorosity is present near the critical point, where the mass-loss rate is set. Such a scenario is supported by several observational
\cite[e.g.][]{puls06, najarro11, bouret12}  and theoretical \citep{cantiello09, sundqvist13} results suggesting that clumping indeed starts close to the photosphere. \\

\begin{figure*}
\centering
\includegraphics[scale=0.41, angle=0]{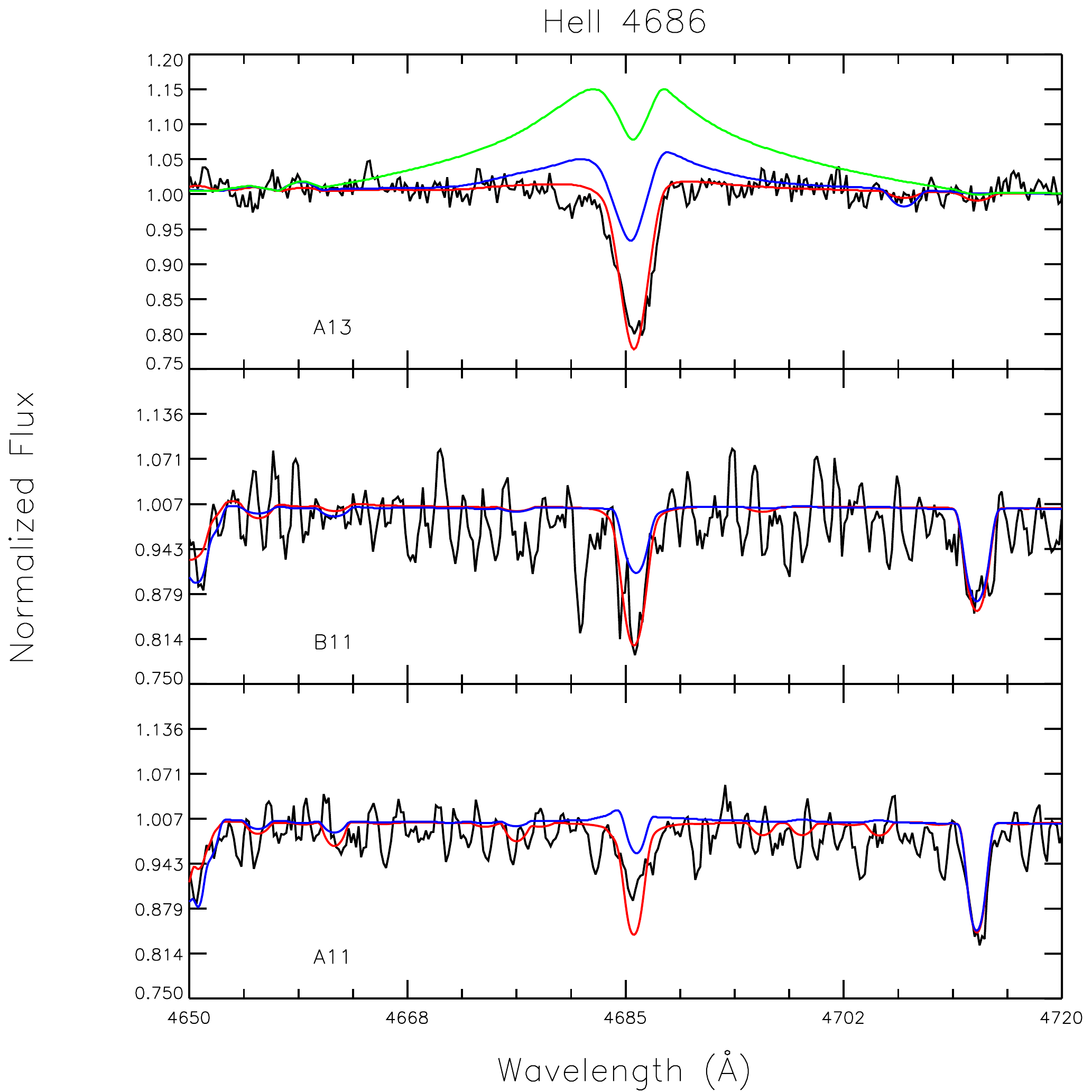}
\includegraphics[scale=0.41, angle=0]{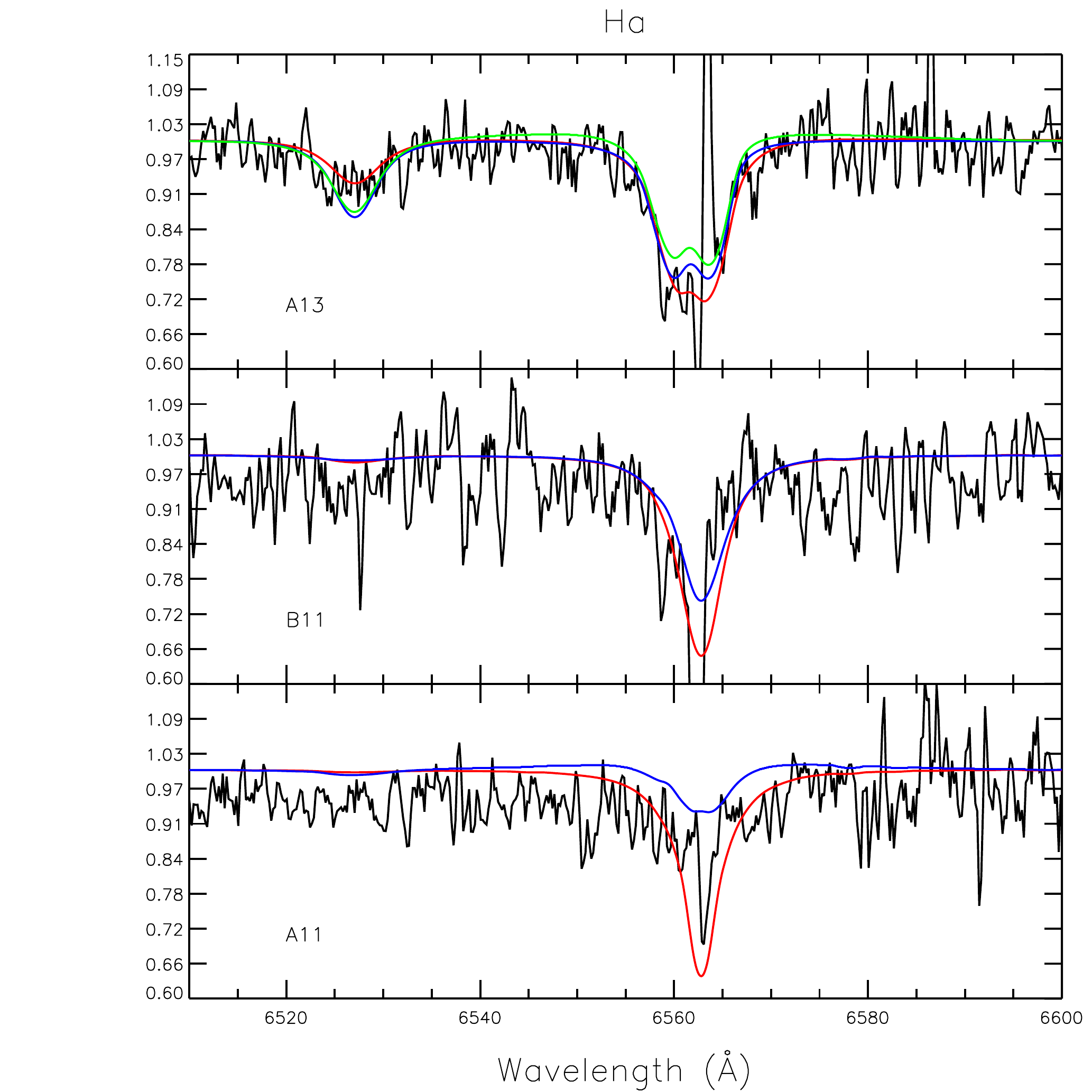}
\caption[]{Left: In black, the spectral region around the \heii\ \lb4686 line, observed for IC1613-A13, B11 and WLM-A11(from top to bottom). Synthetic profiles produced by models 
computed with \cmfgen\ using the parameters from \cite{tramper14} and those derived from the joint FUV+optical analysis presented 
here are overplotted (in blue and red, respectively). Right: Same, from the region around the H$\alpha$ line. The model in green presented for IC1613-A13 illustrates the sensitivity
of the \heii\ \lb4686 line to line-blanketing (\mbox{Ne, Cl, Ar, Ca} are not included) while the H$\alpha$ line is much less affected. 
}
\label{fig8}
\end{figure*}

The mass-loss rates we derived using models with clumping are lower than those obtained from the modeling of the optical spectra only, especially for the late-type supergiants.  
We present in Fig. \ref{fig8} the synthetic profiles of wind sensitive lines in the optical, namely the \heii\ \lb4686 and H$\alpha$ line, as predicted with our values for \mdot\ and those
obtained with \cmfgen\ for the parameters listed in \cite[][their tables 4 and 5]{tramper14}. 

The H$\alpha$ line in IC1613-A13 are very similar in shape and strength, but a striking difference is observed in the predicted profiles for \heii\ \lb4686. Even more striking is the fact this line 
is predicted in absorption by \fastwind\ \cite[][cf. figure 4 in]{tramper14}, while the profile predicted by \cmfgen\ for the same parameters (and no clumping) shows clear electron scattering wings in emission. It is known however that \cmfgen\ often produces more emission in H$\alpha$  and \heii\ \lb4686 than \fastwind, for instance in the temperature and surface gravity regime corresponding to IC1613-A13 \citep[e.g.][]{rivero12}. In this regime, the formation of \heii\ \lb4686 is very sensitive to radiative transfer effects, for instance the detailed treatment and amount of line-blanketing (model
in green in Fig. \ref{fig8} has much less elements than the best fit models), and the line profile responds accordingly. 

Concerning the late-type supergiants IC1613-B11 and WLM-A11, the profiles predicted by \cmfgen\ for \heii\ \lb4686 and H$\alpha$ for the two sets of parameters
bracket the observed profiles. Since, admittedly, \cmfgen\ predicts stronger emission than \fastwind\ in these lines, 
these profiles in Fig. \ref{fig8} also likely bracket those which lead \cite{tramper11, tramper14} to 
claim that mass-loss rates are higher than expected in IC1613 and WLM. We argue that, from such profiles only, it is quite uncertain what \mdot\ should be favored. 
 
Despite the uncertainties on the real absolute values of mass-loss rates introduced by the feedback from clumping on line driving, the considerations spelled out 
hereabove suggest that the empirical
mass-loss rates derived by \cite{tramper14} are likely overestimated, compared to those expected theoretical predictions and those derived by clumped models applied
to the FUV + optical spectral range.

\section{Observed versus predicted mass-loss rates}
\label{sect_pred_wind_prop}
\subsection{Semi-empirical mass-loss rates}
\label{sect_vink}
The mass-loss rates calculated for the stellar parameters using the recipe by \cite{vink01} are presented in Table \ref{tab2}. 
For the three stars, \mdot\ measured from FUV spectroscopy are lower than these estimates. 

In the case of A13, the difference is slightly more than a factor of three, which is reasonable given the uncertainties on the photospheric and wind parameters derived from spectroscopy. 
Note that this difference would be sightly reduced if Z = 0.14Z$_{\odot}$ applies instead to the stars (by a factor $(0.2/0.14)^{0.69-0.83}$ = 1.28--1.34), although the spectroscopic 
analysis does not support this lower value.
The mass-loss rate and clumping parameters of A13  are consistent with recent results obtained for
SMC dwarfs \citep{bouret13}. The same behavior between measured and theoretical mass-loss rates is obtained for early type galactic dwarfs \citep{bouret05, martins12a} and supergiants \citep{bouret12}. 
All in all, this indicates that the wind properties of O stars of early spectral type are well 
accounted for by the standard theory of radiatively driven winds, regardless of the luminosity class and stellar metallicity. 
As for the wind terminal velocity, we note that  our value \vinf\ = 2180 \kms\ is much higher than \vinf\ $\approx$ 1570 \kms, inferred from
the usual scaling \vinf/$v_{esc}$ = 2.6 \citep{kudritzki00}, then to the star's metallicity \vinf\ $\propto Z^{0.13}$ \citep{leitherer92}. It is important to recall that in the recipe by \cite{vink01}, the wind velocity 
structure and terminal velocity are adopted as input to predict the mass-loss rate. Said differently, the adopted value for 
\vinf\ impacts the mass-loss prediction. 
A difference in \vinf\ such as the one we find for A13 implies a difference of more than 0.28 dex in the theoretical mass-loss rates, 
all other parameter being equal. 

The case for B11 and A11 is different. 
The terminal velocities we measure are in good agreement (within 50 \kms) with the values predicted using the theoretical scaling between surface escape velocity and wind terminal velocity, then with metallicity (for the derived stellar parameters and iron abundance).
On the other hand, the mass-loss rates are much lower than predicted by the recipe from \cite{vink01} for these stellar parameters and the iron abundance. 
The discrepancy reaches up to a factor of twenty and thirty five for B11 and A11 (resp.), assuming a SMC-type metallicity for IC 1613 and WLM. 
In order to reconcile the measured mass-loss rates with theoretical values, the metallicities would need to be Z = 0.03Z$_{\odot}$ and 0.002Z$_{\odot}$ for B11 and A11 respectively. 
Such low metallicities are clearly ruled out by the spectroscopic analysis (as is Z = 0.14Z$_{\odot}$, which would only marginally reduce the discrepancy between observed and theoretical
\mdot\ anyway). Of course, this discrepancy is very likely an upper limit, should the true mass-loss rates be revised upward for A11 and B11as a result of a proper treatment of porosity/vorosity
effects (cf. Sect. \ref{sect_error}). Note also that in the specific case of A11, the difference could be reduced though, adopting a luminosity and a mass more in line with those expected for a normal O9.7 supergiant, to compute the theoretical mass-loss rate \citep[e.g. $\log \frac{L}{L_{\odot}}$ = 5.5 and M=31\msun, from][]{martins05}.  When corrected for this luminosity effect, possibly indicating a binary/multiple status for this star, the difference with the theoretical \mdot$_{Vink}$ is of the same amplitude as for B11. 

In any case, even though our modeling concurs with \cite{tramper11, tramper14} that the measured mass-loss rates of B11 and A11 disagree with theoretical predictions, 
\emph{we find the discrepancy to occur in the opposite sense.}

Recently, \cite{muijres12} built on the work by \cite{muller08} to compute new mass-loss rates and terminal wind velocities, improving the physics implemented in the Monte-Carlo (MC) method used by \cite{vink00, vink01}. 
In this new approach, a parametrized description for the line acceleration is used to derive new expression for the velocity field of a radiation-driven wind. A major conclusion of this work is that 
for stars more luminous than $\log \frac{L}{L_{\odot}} \geq\ 5.2$, the new mass-loss rates and those by \cite{vink00} agree well. On the other hand, for physical parameters relevant to a typical 
late-type O supergiant (i.e. close to those of B11 and A11), the new mass-loss rates are between 0.3 -- 0.4 dex lower than those computed by \cite{vink00}.  
The wind terminal velocities however, are notably higher than observed for such objects \citep[e.g.][]{howarth89}. Notwithstanding the possible reasons for these over-predicted \vinf\ \cite[see][for a discussion]{muijres12}, these MC simulations essentially predict the total wind kinetic energy \mdot\vinf$^{2}$, and any change in the wind terminal speed reflects in a change in the mass-loss rate, 
hence the observed decrease of \mdot. More quantitative comparison to B11 and A11 is not possible, however, because the mass-loss rates and wind terminal velocities have been computed by \cite{muijres12} for galactic metallicity only. Assuming that the theoretical scaling obtained by \cite{vink01} still holds, the mass-loss rate should be scaled down by another 0.5 dex, while \vinf\ would change by less than 20\%, compared to the galactic values. 
With a total scale down of $\approx$ 0.8 -- 0.9 dex, \mdot\ would still be a factor of three to five higher than measured from FUV spectroscopy. Furthermore, if the \vinf\ for these late-type supergiants are much 
lower (as we measure them) than predicted (40\% typically), then the theoretical \mdot\ should go up. It is not clear that mass-loss prediction at different metallicities, based on \cite{muijres12} improvement in the treatment of the line force parametrization, would be able to account for the observed mass-loss rates of B11 and A11.

\subsection{Mass fluxes obtained by solving the dynamics of radiatively-driven winds}
\label{sect_lucy}
\cite{lucy10a} argued that the mass-loss rates by \cite{vink01} that have been compared to observations, may not be relevant for such comparisons 
because they are derived by imposing a global dynamical constraint, i.e. assuming that the energy that is needed to drive the wind is indeed extracted from the radiation field,  rather than being obtained
by solving the equations governing the dynamics of radiatively-driven winds. In order to fully guarantee that the derived mass-loss rates are consistent with stationary transonic flows, \cite{lucy10a}
computed mass fluxes that allow the wind mass flow to accelerate continuously from sub- to supersonic velocities. The theory of moving reversing layers (MRL) pioneered by \cite{lucy70} was used to determine
the mass fluxes from first principles by imposing a regularity condition at the sonic point.
An important conclusion from this work was that the so-called \it weak-wind problem\rm\ mostly arises from the existence of two separate regimes of ionization of iron. 
More precisely,
a shift in the ionization balance of iron, from \fev\ to \feiv\ (as \teff\ decreases), leads to weaker line acceleration below the sonic point, which translates into lower mass-loss rates. 
In the temperature domain of late O-type stars, namely \teff\ $\approx$ 30,000~K, the \fev\ contribution to the driving is dramatically reduced and a reduction by up to $\approx$ 1.4 dex is 
obtained compared to mass-loss rates predicted by \cite{vink00}\footnote{\cite{muijres12} do solve the full momentum equation, but they now obtain that for stars
having  $\log \frac{L}{L_{\odot}} \geq\ 5.2$, i.e. stars in the ``\it weak wind\rm'' regime, their hydrodynamical method fails to drive a wind.}. 

More recently, \cite{lucy12} extended this work to predict mass fluxes that depend on the stellar metallicity, without relying on scaling laws \citep[e.g.][]{vink01, mokiem07}
but implementing curve-of-growth effects in the dynamic of the reversing layers instead. 
Mass fluxes computed with this method present manifest departure from the expected, linear, monotonic, decline with metallicity \citep[see Fig. 1 in][]{lucy12}. 
It is noteworthy that the strongest departure from a (single) power-law scaling is found for the metallicity and (\teff\ -- \logg) regime of the two late-type supergiants IC 1613-B11 and WLM-A11. 
 
We used the value listed in Table 1 from \cite{lucy12} to calculate the mass-loss rates corresponding to the stellar parameters we derive for IC1613-A13, IC1613-B11 and WLM-A11 respectively. 
It is fortunate that the case for \teff = 42,500 K and \logg = 3.75 was also worked out by \cite{lucy12} as it best account for the parameters of IC1613-A13 as derived in the present study. 
The results are summarized in Table \ref{tab2}. The striking conclusion from this comparison is that, while the good match between theoretical and observed \mdot\ for IC 1613-A13 is maintained,
the discrepancy between observed and theoretical mass-loss rates we found for IC1613-B11 and WLM-A11 when using mass-loss rates from \cite{vink01} is dramatically reduced 
(and would be even more if mass-loss rates are higher by a factor of 3 to 10, as discussed in Sect .\ref{sect_error}). 
Quantitatively, the theoretical \mdot$_{Lucy}$ of B11 and A11 are more than one order of magnitude less than \mdot$_{Vink}$. For a large part, this is a consequence of the aforementioned reduction of line-driving 
caused by the lower ionization of iron. The amplitude of the decrease is an additional balance between the decreased contribution of Fe (and Ni for that matter) to the driving and an
enhancement of \mdot\ caused by the lower surface gravities \cite[see Sect. 4 and 5 in][for a discussion]{lucy10b}. 
The remaining differences are well within the uncertainties caused by the cumulative effects of error on the measured stellar and wind parameters (especially the impact of
clumping on the driving) on one hand, and  
in input abundances (global scaling to solar values) and micro turbulence velocity from the \tlusty\ models of \cite{lanz03} used by \cite{lucy12} to compute mass fluxes.

Interestingly, \cite{lucy12} found no dynamically consistent solutions that would yield higher mass fluxes, within the purely radiation-driven winds framework, to provide better agreement with the mass-loss rates by \cite{tramper11}. An (yet unknown) additional driving mechanism would be therefore required to provide more radiative acceleration. With \mdot\ derived from FUV spectroscopy,
such a putative mechanism may not be needed anymore.

\subsection{Modified Wind-Momentum-Luminosity Relation}
\label{sect_wlr}
\begin{figure}
\includegraphics[scale=0.45, angle=0]{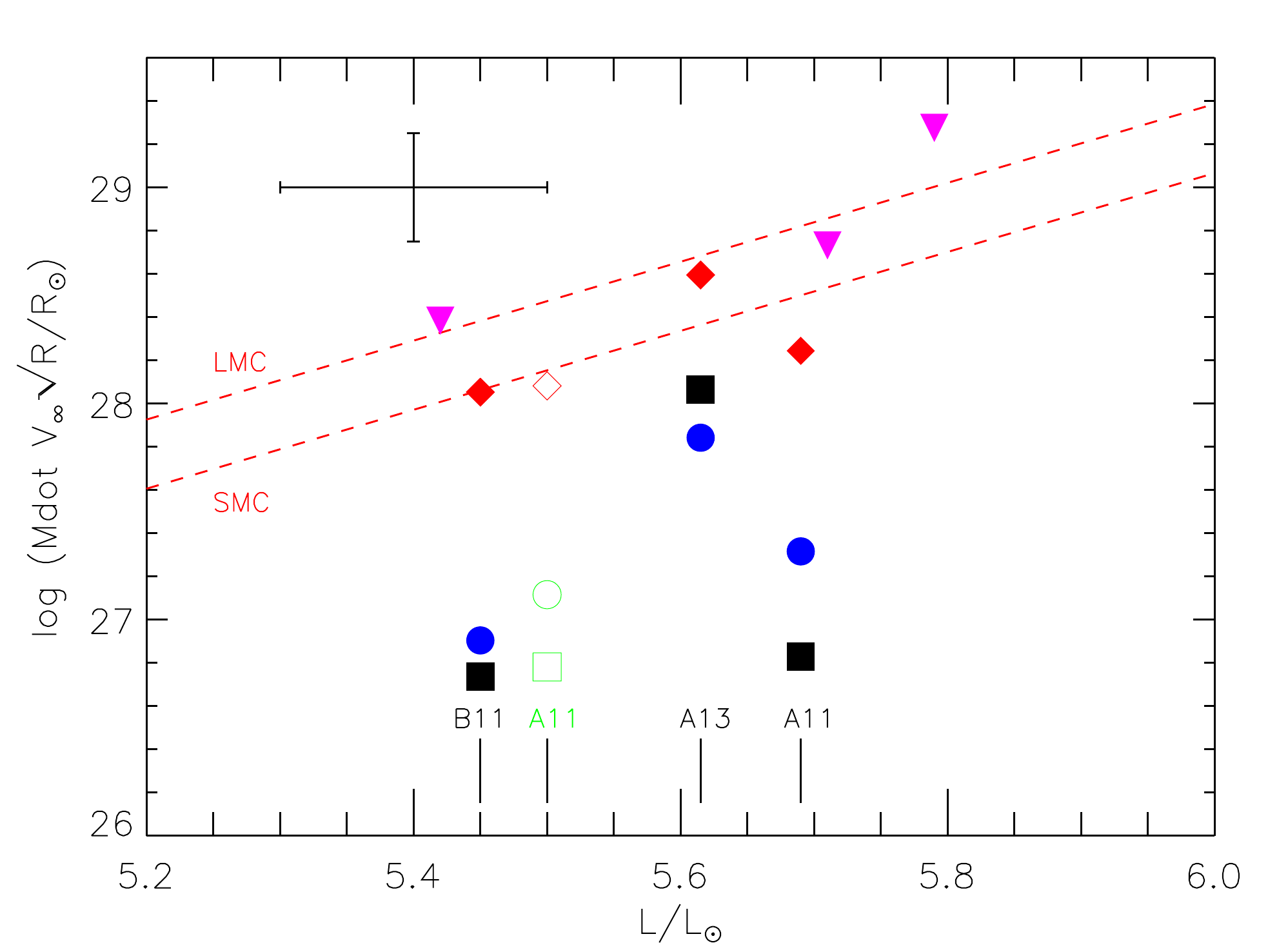}
\caption[]{Modified wind-momentum luminosity relation for the sample stars. Black full squares mark \dmom\ calculated for the spectroscopic parameters listed in Table \ref{tab2}. 
Blue full circles are \dmom\ calculated with the mass fluxes from \cite{lucy12} for the stellar parameters, while red full diamonds are \dmom\ computed with the \cite{vink01} recipe. 
\dmom\ for WLM-A11 assuming \logL\ = 5.5 are indicated as empty symbols (same shape and color coding 
as previously). The dashed red lines are the theoretical relation for LMC and SMC metallicities predicted by \cite{vink00}.
Pink triangles are \dmom\ from \cite{tramper14} for the three stars. 
For clarity, no individual error bars are given but we plot a typical error bar on the upper left corner. 
}
\label{fig9}
\end{figure}

Using the FUV-measured mass-loss, and those computed with \cite{vink01} and \cite{lucy12}, we built the modified wind-momentum-luminosity \citep[see e.g.][]{kudritzki00} relation for the three stars (Fig. \ref{fig9}). 
We further added the modified wind-momenta (\dmom\, cf. Sect. \ref{sect_obs_wind_prop}) obtained for mass-loss rates predicted by both approaches for A11 assuming a lower luminosity 
(hence radius) for this object. Recall that this assumption is based on considerations on measured surface gravity, which yield abnormally high spectroscopic mass for this object,
if the luminosity deduced from the the SED fitting of observed fluxes is used. 
Measured wind-momenta are clearly lower (more than one order of magnitude for B11 and A11) than predicted using the theoretical  coefficients \citep{vink01} for the WLR of the LMC and SMC metallicities (see caption for details). 
For the stellar coefficients derived from spectroscopy, the mass-loss rates predicted by \cite{vink01} yield \dmom\ that are (expectedly) in better agreement with the theoretical relations, although some
striking differences are observed. Note the case of IC1613-A13, where the predicted \dmom\ is more compatible with the theoretical WLR for LMC. The predicted 
\dmom\ for WLM-A11 is below the theoretical relation but the agreement would be much improved assuming that WLM-A11 has a \logL\ lower by 0.2 dex. 

Although the small number of points makes further quantitative interpretation questionable,  we argue that wind properties derived from FUV diagnostics are better accounted for by the MRL approach.
The above comparisons strongly suggest that the information on wind quantities available in the optical spectra in \cite{tramper11, tramper14} is probably too limited and the FUV provides more realistic estimates, while a proper treatment of  the radiatively-driven winds dynamics considerably improves the calculation of mass fluxes. 
The remaining discrepancies between the observed and theoretical mass-loss rates for A13, B11 and A11 reflects the many uncertainties related to
simplifications that are made both in the diagnostic models and in theoretical calculation of mass-loss rates. The mass-loss -- metallicity scaling breakdown at sub-SMC metallicity is not confirmed at this step.

\section{Summary and conclusions}
\label{sect_conclusion}
We have modeled the FUV spectra of three O stars in the low metallicity galaxies IC 1613 and WLM,  observed with HST/\cosp. Our goal was to coroborate or refute the results by \cite{tramper11, tramper14}
suggesting that the mass-loss rates of some of these stars are more compatible with higher metallicity and similar to those of LMC stars. 
This result, if upheld, would have far-reaching consequences beyond our understanding of radiatively-driven winds, and
would impact for instance the predicted number of collapsars and of SN Ib and Ic in low metallicity environments,
hence in the high-z Universe.

Because of numerous iron lines and lines sensitive to wind properties, FUV spectroscopy is the adequate tool to fully address and resolve this outstanding issue of the dependence of hot,
massive star mass-loss rates with metallicity. A comparison of these new \cosp\ spectra with existant \stis\ or \cosp\ spectra of Galactic, LMC and SMC stars provided a direct, model-independent check of the mass-loss - metallicity relation. A quantitative analysis was then carried out using state-of-the-art NLTE unified model atmospheres calculated with the \cmfgen\ code to establish the wind properties of the three low-metallicity massive stars.

\begin{itemize}
\item The spectroscopic analysis indicates that the metallicity of the three targets is 1/5\zsun, i.e. SMC-like, although the usual value adopted for IC1613 and WLM of Z = 1/7\zsun\ is not ruled out. Lower and higher values of the metal content (1/10\zsun\ and 1/2\zsun\ respectively) are not consistent with the observed spectra.
\item The FUV data allow secure inferences on the \mbox{CNO} abundances, of critical importance for mass-loss rates determinations.
\item The wind parameters derived from the FUV-spectroscopic analysis of the observed HST/\cosp\ spectra provide very satisfactory fits to the optical spectra as well. 
This further illustrates the limitations related to \mdot\ determinations based on the optical spectrum (H$\alpha$) only. 
\item The mass-loss rates differ from those derived by \cite{tramper11, tramper14}. The difference for IC1613-A13 is moderate (less than a factor of 6), and it
can be accounted for by clumping (and to some extent to the detailed treatment and amount of line-blanketing). 
For the late-type supergiants (IC1613-B11 and WLM-A11), the amplitude of the difference is much higher, reaching 
from $\sim$ 1.6 to 2.5 orders of magnitude, respectively. Here, wind clumping (absent in the analysis of the optical spectra) might account for a factor of three in the 
difference only. A combination of several other factors can explain the rest of the difference : $(i)$  In these weak wind stars, the influence of porosity/vorosity effects, 
and the possible presence of hot gas, can potentially bias the estimated mass-loss rate by factors from 3 to 10; $(ii)$ In the region of parameter space corresponding to
these stars, \cmfgen\ and \fastwind\ produce systematically different results for optical lines, with more emission/infill predicted by \cmfgen.
\item For the re-visited metallicity, we find that the mass-loss rates of IC1613-A13, the earliest star of the sample, is in good agreement with the mass-loss predicted with the recipe by \cite{vink01}.  The \civ\ P~Cygni profile of A13, which is stronger than that of the SMC dwarf MPG 324 (cf. Fig. \ref{fig2}) is probably a consequence of its more evolved status.
\item On the other hand, \mdot\ measured with FUV spectra for IC1613-B11 and WLM-A11, two late-type O supergiants, are clearly lower than measured in the optical by \cite{tramper14}. 
Because of  the systematic errors, they may however be consistent with the theoretical predictions by \cite{vink01} for the preferred metallicities.  
If the later is true, porosity/vorosity effects  and the state of the gas (i.e., what fraction is hot)  are of crucial importance.
%
\item We used a different approach to calculate mass-loss rates, as outlined recently in \cite{lucy12}. With this alternative method, avoiding scaling laws and assumptions on the terminal wind velocities,
the measured mass-loss rates of the three stars are then in much better agreement with theoretical values. We therefore conclude that there is no indication at this moment that the radiatively-driven wind theory breaks down at metallicities at or slightly less than SMC values. 
\end{itemize}

Our results are restricted to a very limed number of objects. 
We pointed out that the observed mass-loss rates within a spectral/luminosity class may vary within a factor
of four or more just because of the scatter of the intrinsic individual parameters. To reach a firm conclusion on the nature of the \mdot(Z) relation therefore requires larger statistical sample, preferentially 
observed in the FUV at (relatively) high spectral resolution and high signal-to-noise ratio, to simultaneously derive the iron content and the wind parameters. 
This, unfortunately, is beyond routine capability of HST for galaxies more distant than the MCs. 
On the other hand, a fairly decent sample of objects is available in the HST archives for stars in the SMC and we shall now focus on these objects to study the properties of stellar winds at low metallicity. 
We engaged in such a study and results will be presented in a coming paper discussing the properties of mass-loss for stars from dwarfs to supergiants.

\label{Sec_future}

\section*{acknowledgements}
We thank an anonyous referee for a careful reading and suggestions which help improve this paper.
We thank Joachim Puls for computing a \fastwind\ model for us and for discussion about wind sensitive optical lines.  
J.-C. Bouret is indebted to S. R. Heap for her invitation to work at NASA/GSFC when
this work was initiated (work was supported by NASA grant NNX08AC146
to the University of Colorado at Boulder).
DJH acknowledges support from STScI GO grant HST-GO- 12867.001-A and STScI theory grant HST-AR- 12640.01.
We acknowledge financial support from ``Programme National de Physique
Stellaire'' (PNPS) of CNRS/INSU, France.
This research has made use of the SIMBAD database, operated at CDS, Strasbourg, France. 

\bibliography{biblio_lowz}

\begin{appendix}
\label{appendice}

\section{Best fits}
In this appendix we compare our best-fit models to the optical and UV spectra.
Absolute fluxes in UV spectra are provided after correcting for reddening (see Sect. \ref{sect_model}) and are 
expressed in ergs\,.\,\AA$^{-1}$.\,s$^{-1}$.\,cm$^{-2}$.  

 \begin{figure*}
\centering
\includegraphics[scale=0.8, angle=0]{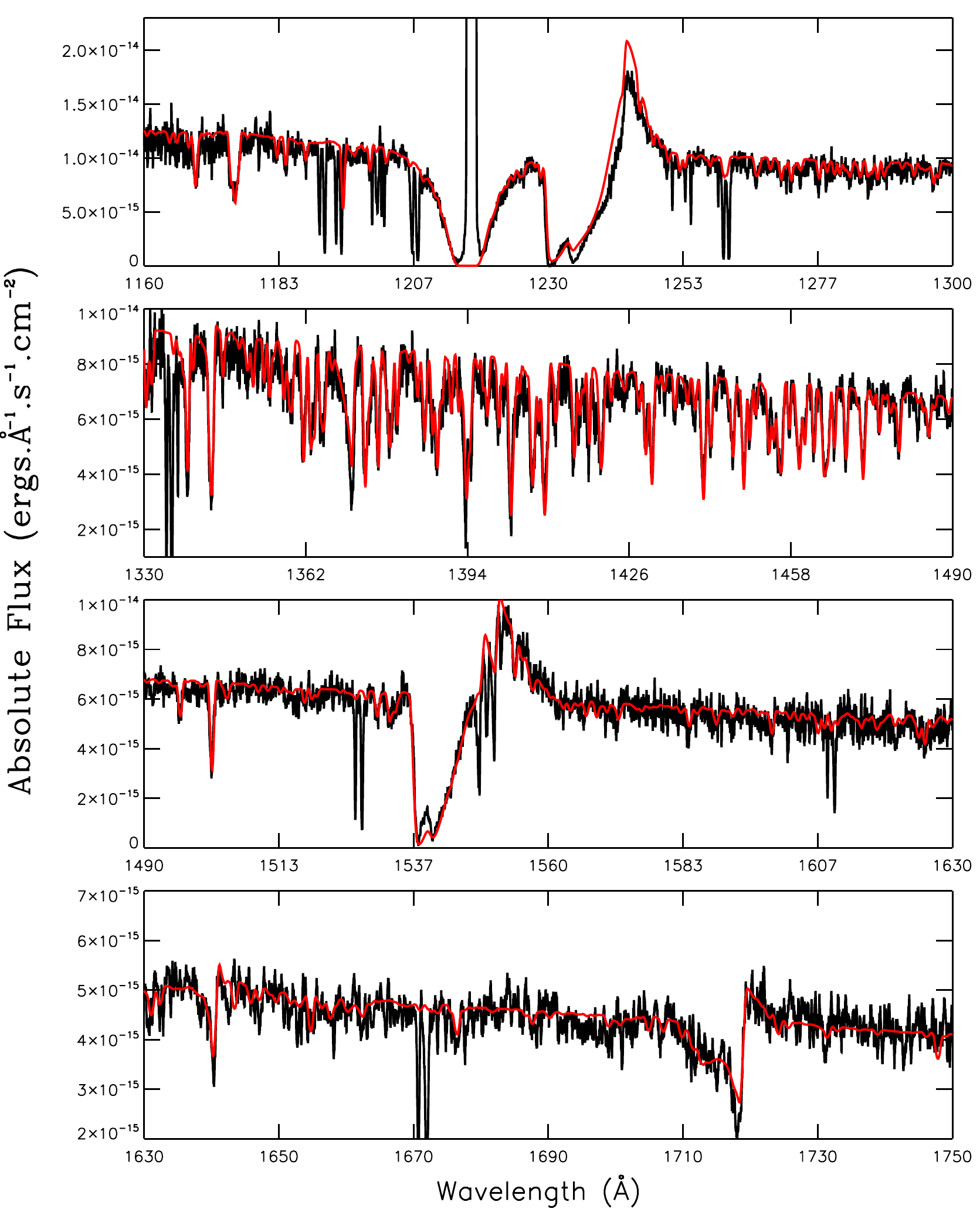}
\caption[]{FUV spectrum of IC 1613-A13 (in black). Overplotted in red is the best fit model (see text for comments). A satisfactory fit is obtained for the observed photospheric (mainly \fe\ forest) and wind
profiles. In particular, the \nv\ \lb1240 and \niv\ \lb1718 wind profiles are well reproduced, indicating correct ionization factions in the wind.}
\label{fig_a1}
\end{figure*}

\begin{figure*}
\centering
\includegraphics[scale=0.8, angle=0]{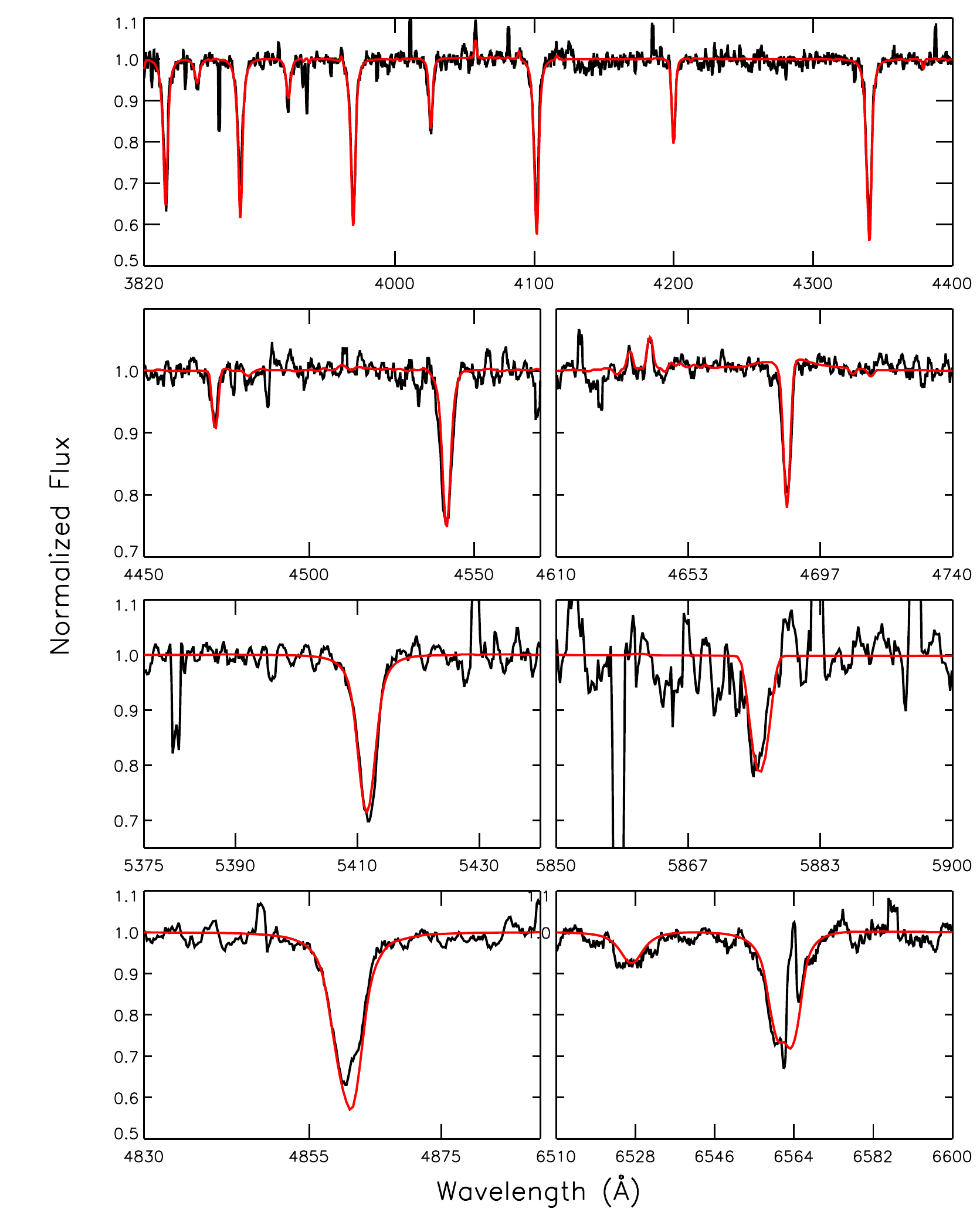}
\caption[]{Optical spectrum of IC 1613-A13. Overplotted in red  is the best fit model (see text for comments).}
\label{fig_a2}
\end{figure*}

 \begin{figure*}
\centering
\includegraphics[scale=0.8, angle=0]{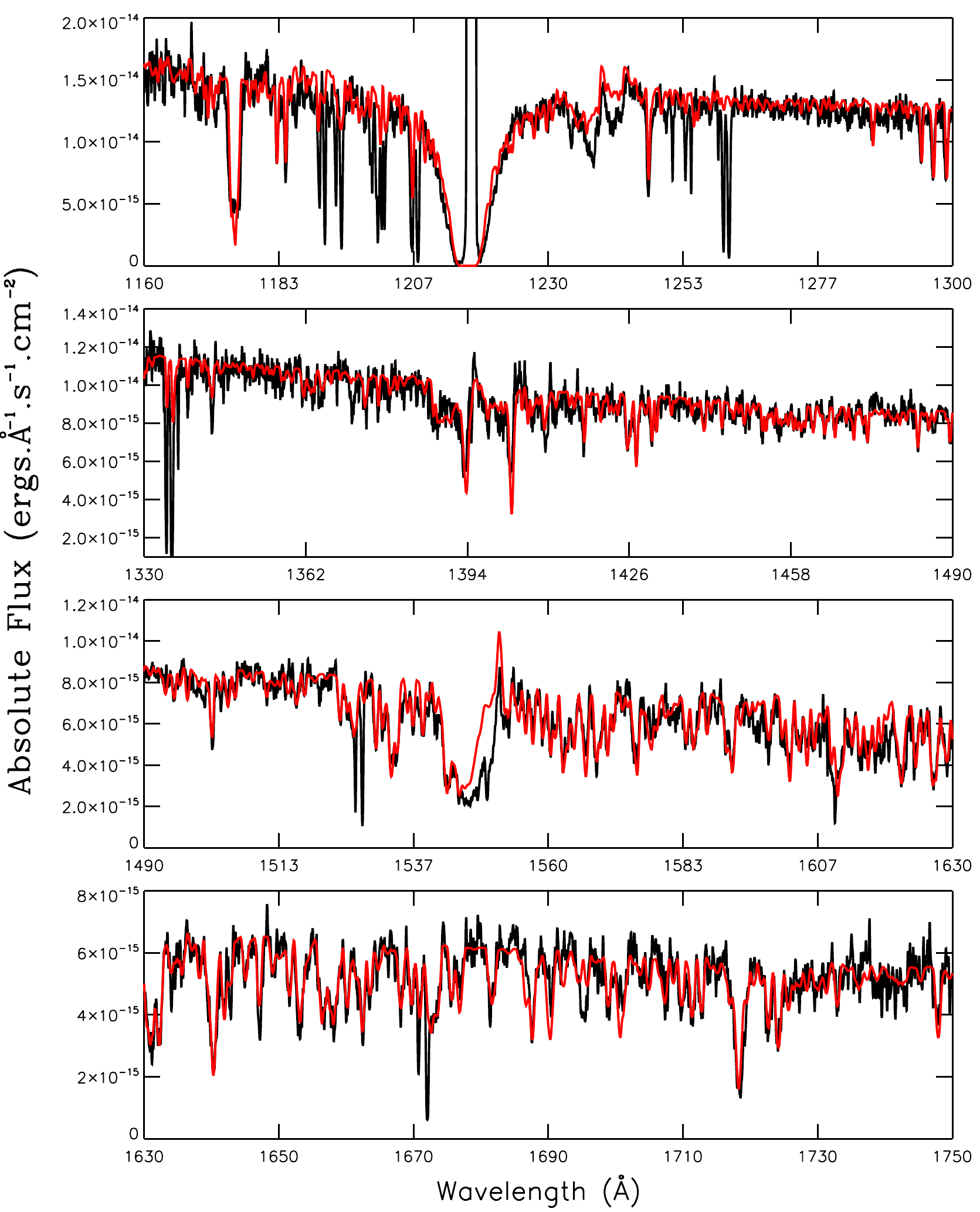}
\caption[]{FUV spectrum of IC 1613-B11 (in black). Overplotted in red is the best fit model (see text for comments).}
\label{fig_a3}
\end{figure*}

\begin{figure*}
\centering
\includegraphics[scale=0.8, angle=0]{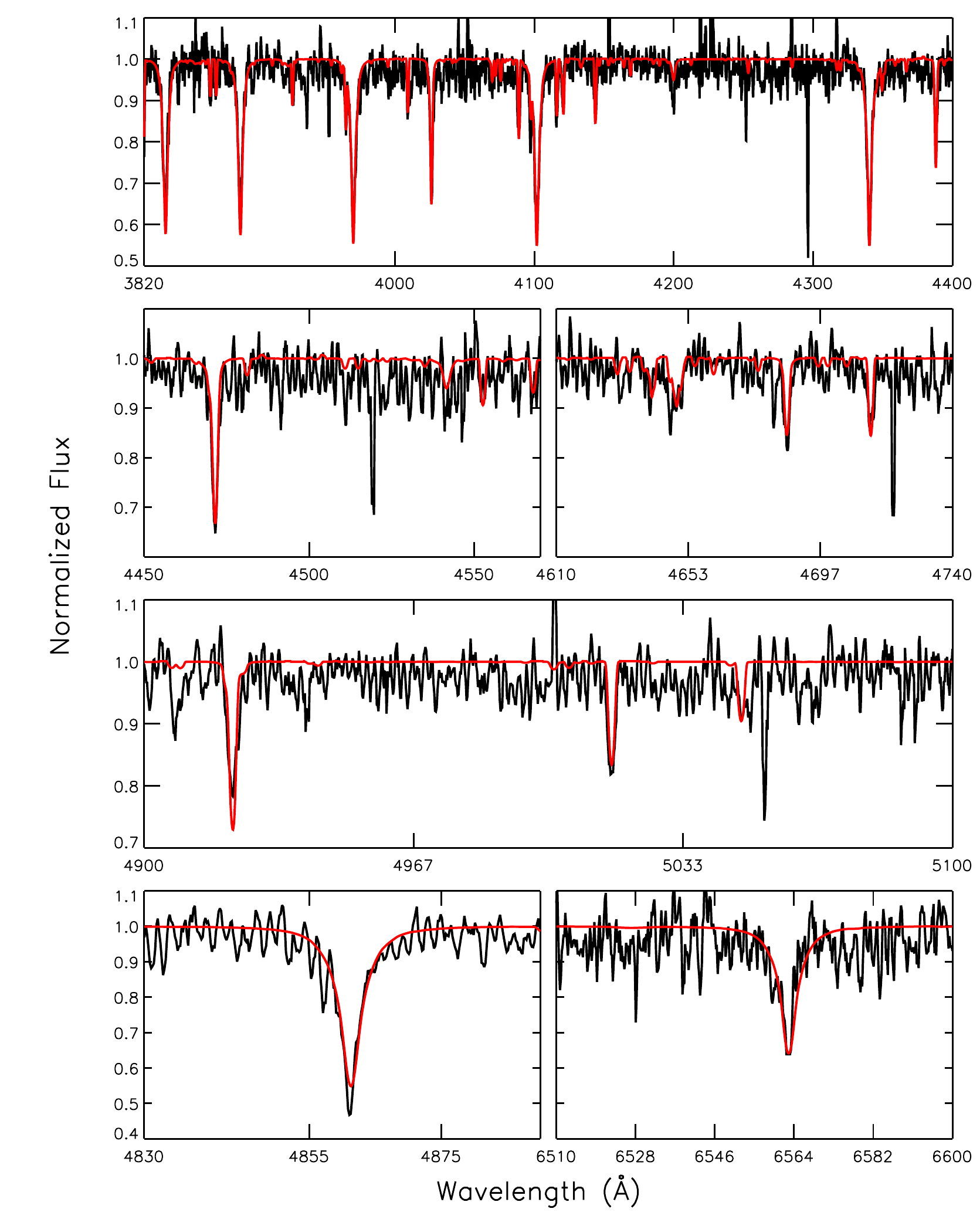}
\caption[]{Optical spectrum of IC 1613-B11. Overplotted in red  is the best fit model (see text for comments).}
\label{fig_a4}
\end{figure*}

 \begin{figure*}
\centering
\includegraphics[scale=0.8, angle=0]{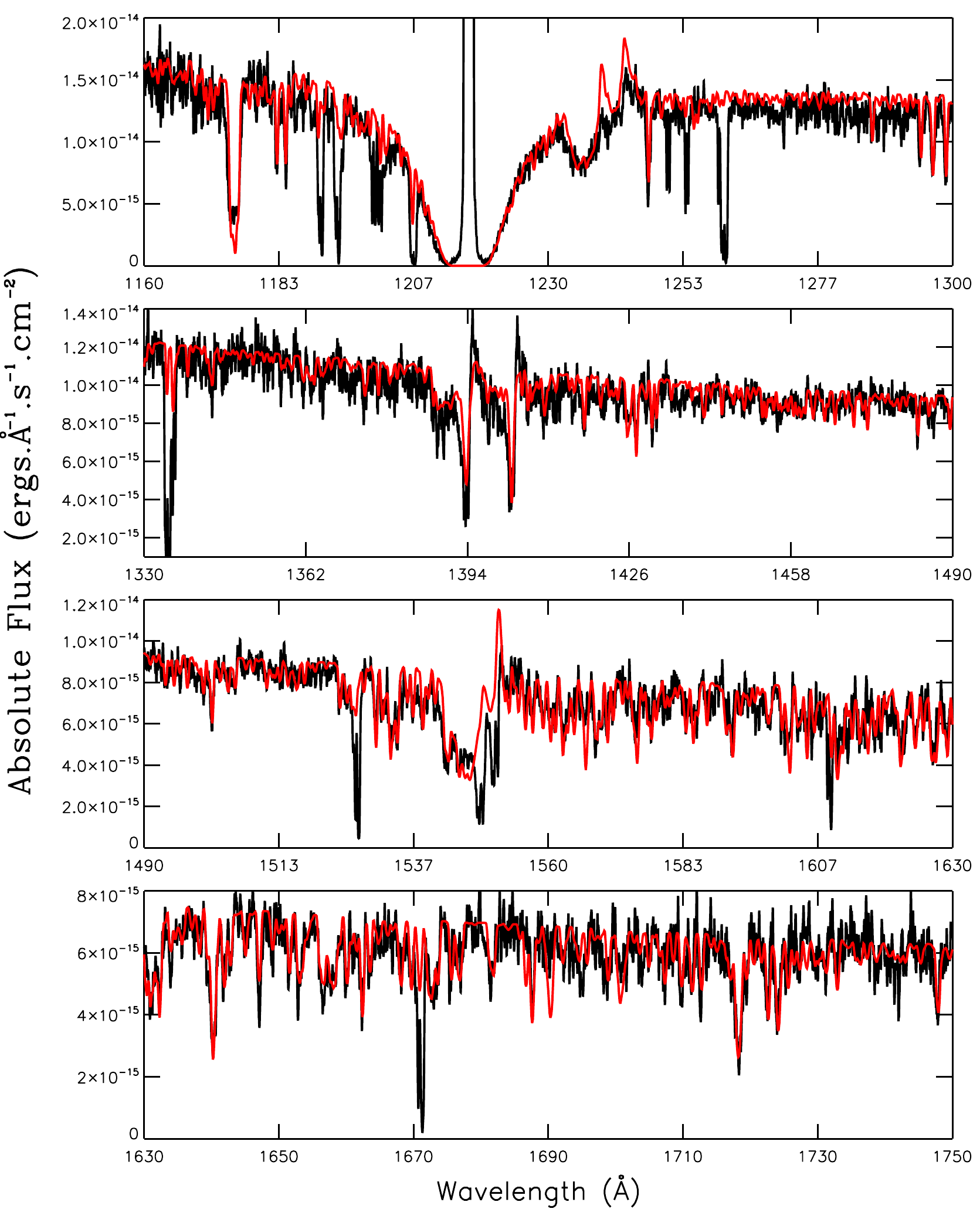}
\caption[]{FUV spectrum of IC WLM-A11 (in black). Overplotted in red is the best fit model (see text for comments).}
\label{fig_a5}
\end{figure*}

\begin{figure*}
\centering
\includegraphics[scale=0.8, angle=0]{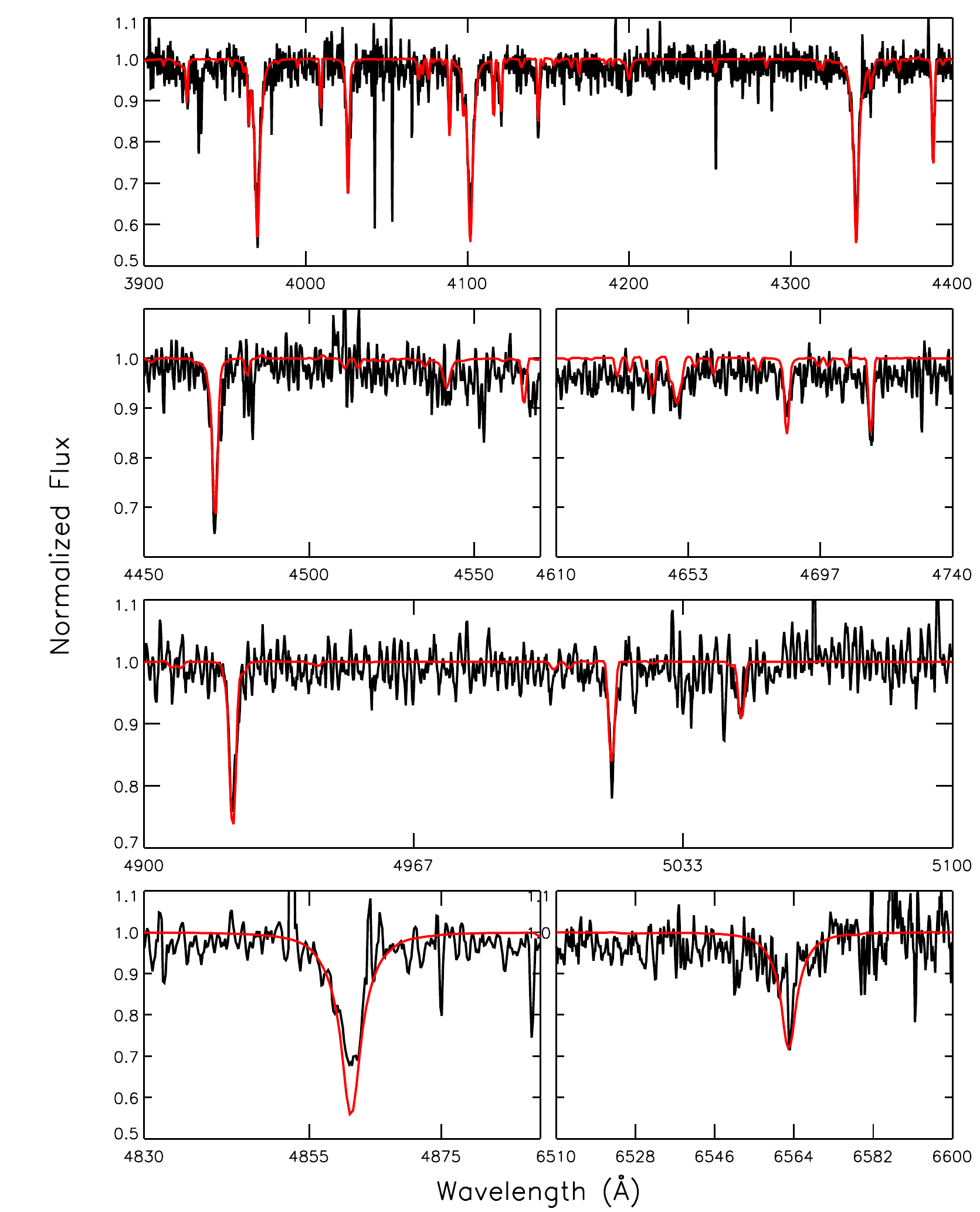}
\caption[]{Optical spectrum of WLM-A11. Overplotted in red  is the best fit model (see text for comments).}
\label{fig_a6}
\end{figure*}

\end{appendix}
\label{lastpage}
\end{document}